\documentclass[amssymb,amsmath,prd,twocolumn,superscriptaddress,nofootinbib]{revtex4-1}

\usepackage{graphicx}
\usepackage{bm}
\usepackage{amssymb,amsmath}
\usepackage{mathrsfs}
\usepackage{latexsym}
\usepackage{color}
\usepackage[normalem]{ulem}
\usepackage{dcolumn}
\usepackage[colorlinks=true,citecolor=blue,urlcolor=blue]{hyperref}
\usepackage[usenames,dvipsnames]{xcolor}
\usepackage{xspace}
\usepackage{graphicx}

\allowdisplaybreaks

\newcommand{\MSFC}{\affiliation{NASA Marshall Space Flight Center, Huntsville, AL 35812, USA}}
\newcommand{\XGI}{\affiliation{eXtreme Gravity Institute, Department of Physics, Montana State University, Bozeman, Montana 59717, USA}}
\newcommand{\CIT}{\affiliation{Department of Physics, California Institute of Technology, Pasadena, California 91125, USA}}
\newcommand{\CITLab}{\affiliation{LIGO Laboratory, California Institute of Technology, Pasadena, California 91125, USA}}
\newcommand{\US}{\affiliation{Mathematical Sciences and STAG Research Centre, University of Southampton, SO17 1BJ, Southampton, UK}}

\definecolor{kcmagenta}{rgb}{0.54, 0.17, 0.88}
\definecolor{shyellow}{rgb}{0.15625, 0.609375, 0.316406}
\definecolor{chorange}{rgb}{0.851, 0.372, 0.007}
\definecolor{tlteal}{rgb}{0,.55,.55}
\definecolor{jcpink}{rgb}{1.0, 0.0, 0.5}
\definecolor{mmgreen}{rgb}{0.0, 0.8, 0.6}
\definecolor{bbsalmon}{rgb}{1.0, 0.47, 0.42}

\newcommand{\BayesWave}{{\tt BayesWave}\xspace}

\newcommand{\chieff}{\chi_{\textrm{eff}}}

\newcommand{\comment}[1]{}
\newcommand{\Tobs}{T_{\textrm{obs}}}
\newcommand{\ttrig}{t_{\textrm{trig}}}
\newcommand{\Tw}{T_{\textrm{w}}}
\newcommand{\chip}{\chi_{\textrm{p}}}
\newcommand{\Qmax}{Q_{\textrm{max}}}
\newcommand{\Dmax}{D_{\textrm{max}}}
\newcommand{\flow}{f_{\textrm{low}}}

\newcommand{\gCBC}{g_{\textrm{CBC+G}}} 
\newcommand{\cbcG}{h_\textrm{CBC+G}^\textrm{rec}} 
\newcommand{\gG}{g_{\textrm{G}}} 
\newcommand{\inj}{h^\textrm{inj}} 
\newcommand{\cbcCBC}{h_\textrm{CBC}^\textrm{rec}}  
\newcommand{\OgGinj}{ {\mathcal{O}}(\gG | \inj)} 
\newcommand{\OgCBCgG}{{\mathcal{O}}(\gG | \cbcG)} 
\newcommand{\MgGgCBC}{ {\mathcal{M}}(\gG | \gCBC)} 
\newcommand{\McbcGinj}{{\mathcal{M}}(\cbcG | \inj)} 
\newcommand{\McbcCBCinj}{{\mathcal{M}}(\cbcCBC | \inj)} 
\newcommand{\OcbcCBCgG}{{\mathcal{O}}(\cbcCBC | \gG)} 

\begin{document}
\title{Accurate modeling and mitigation of overlapping signals and glitches in gravitational-wave data}

\author{Sophie Hourihane} \CIT \CITLab 
\author{Katerina Chatziioannou} \CIT \CITLab 
\author{Marcella Wijngaarden} \US
\author{Derek Davis} \CITLab 
\author{Tyson Littenberg} \MSFC
\author{Neil Cornish} \XGI

\date{\today}

\begin{abstract}
The increasing sensitivity of gravitational-wave detectors has brought about an increase in the rate of astrophysical signal detections as well as the rate of ``glitches"; transient and non-Gaussian detector noise. Temporal overlap of signals and glitches in the detector presents a challenge for inference analyses that typically assume the presence of only Gaussian detector noise. In this study we perform an extensive exploration of the efficacy of a recently proposed method that models the glitch with sine-Gaussian wavelets while simultaneously modeling the signal with compact-binary waveform templates. We explore a wide range of glitch families and signal morphologies and demonstrate that the joint modeling of glitches and signals (with wavelets and templates respectively) can reliably separate the two. We find that the glitches that most affect parameter estimation are also the glitches that are well modeled by such wavelets due to their compact time-frequency signature. As a further test, we investigate the robustness of this analysis against waveform systematics like those arising from the exclusion of higher-order modes and spin-precession effects.
Our analysis provides an estimate of the signal parameters; the glitch waveform to be subtracted from the data; and an assessment of whether some detected excess power consists of a glitch, signal, or both. We analyze the low-significance triggers (191225\_215715 and 200114\_020818) and find that they are both consistent with glitches overlapping high-mass signals.  
\end{abstract}

\maketitle

\section{Introduction}\label{sec:intro}

Gravitational-wave (GW) analyses require accurate models for both the astrophysical signals and the detector noise~\cite{LIGOScientific:2019hgc}.
The majority of source properties inference for transient signals such as compact binary coalescences (CBCs) is based on three assumptions about the detector noise that inform the functional form of the likelihood function: (i) the detector noise is uncorrelated between the detectors, (ii) it follows a Gaussian distribution, and (iii) it is stationary, i.e. its mean and covariance do not change with time. Violation of these assumptions could impact detection and inference efforts. For example, Schumann resonances in the Earth's large-scale magnetic field could cause correlated detector noise and affect detection and interpretation of a stochastic GW background~\cite{Thrane:2013npa,Thrane:2014yza,Himemoto:2017gnw,Himemoto:2019iwd,Callister:2018ogx,Coughlin:2018tjc,Meyers:2020qrb}. Additionally, the detector Gaussian noise is stationary over short timescales~\cite{Chatziioannou:2019}, but longer signals might be subject to noise nonstationarity which has motivated relevant studies~\cite{Cornish:2020odn,2021PhRvD.103l4061E,Mozzon:2021wam,Vajente:2019ycy}.

Transient noise artifacts, i.e., glitches, in a detector violate the assumption of Gaussianity and could bias parameter inference when they overlap with signals~\cite{Powell:2018csz,Pankow:2018qpo, Driggers:2018gii}. In the recent third observing run (O3), LIGO~\cite{TheLIGOScientific:2014jea} and Virgo~\cite{TheVirgo:2014hva, VIRGO:2014yos} have detected an astrophysical event approximately once every five days~\cite{gwtc3, O3a_catalogue}. Glitches, however, appear in the detectors far more frequently. The average rates for glitch transients with signal-to-noise ratio $(\text{SNR}) > 6.5$ in the first and second half of the third observing run were $0.3\,\text{min}^{-1}$ in LIGO Hanford (LHO), $1.13\, \text{min}^{-1}$ in LIGO Livingston (LLO), and $ 0.75 \, \text{min}^{-1}$ in the Virgo detector~\cite{Abbott:2020niy,gwtc3}. Overall, in O3 a total of 18 events required some form of glitch mitigation~\cite{Abbott:2020niy,gwtc3}. Glitches are most likely to intersect lower-mass, long-duration events such as binary neutron star (BNS) mergers; indeed, both such detected events have overlapped with a glitch and required mitigation~\cite{TheLIGOScientific:2017qsa, GW190425}. As detector sensitivity improves not only will the event rate increase, but also the glitch rate might increase as weaker glitches that are currently below the noise floor could emerge above it. Additional detectors such as KAGRA~\cite{KAGRA:2018plz} in the next observing run (O4) also increase the likelihood that a glitch will appear in at least one detector. In order to have an accurate and unbiased catalog of GW events, effective and generic methods for separating signal and glitch power are necessary. Proposed approaches for glitch subtraction include removing the contaminated data~\cite{TheLIGOScientific:2017qsa,Usman:2015kfa,Sachdev:2019vvd,Zackay:2019kkv,Steltner:2021qjy, Zweizig:gating} or subtracting detector noise based on data from auxiliary channels~\cite{mogushi2021reduction,Tiwari:2015ofa,Meadors:2013lja,LIGOScientific:2018kdd,Davis:2018yrz,Ormiston:2020ele,Yu:2021swq,Colgan:2022vdd,Vajente:2019ycy, Viets:narrowband}. The glitches discussed in this paper are those that remain after the noise mitigation described in~\cite{Davis:2018yrz, Viets:narrowband}. 

A complementary analysis was proposed in Ref.~\cite{Chatziioannou:2021ezd} based on the \BayesWave algorithm~\cite{Cornish:2014kda,Littenberg:2014oda,Cornish:2020dwh}. This analysis expands glitch-mitigation techniques already applied to LIGO and Virgo data~\cite{TheLIGOScientific:2017qsa,Abbott:2020niy,gwtc3}, where it was used to subtract glitches in 15 out of 18 O3 events that required glitch mitigation~\cite{Abbott:2020niy,gwtc3}. The analysis of~\cite{Chatziioannou:2021ezd} simultaneously models the signal and glitch using waveform templates and wavelets respectively. The waveform templates are models for CBC signals that are obtained as solutions to the 2-body problem in General Relativity. The glitch model is based on a sum of sine-Gaussian wavelets and it is flexible enough to be able to reliably describe a wide range of potential glitch morphologies. This first study presented a number of examples of data containing binary black hole (BBH) signals and common glitch types and demonstrated the ability of the analysis to reliably separate the two~\cite{Chatziioannou:2021ezd}. In this study we expand upon this analysis by considering a wider rage of glitch classes, more instances of each class, and CBC injections of varying masses. 

Our analysis results in a posterior distribution for the parameters of the glitch and the CBC signal. Depending on the exact placement of the signal in relation to the glitch, correlations between the two might exist and the resulting CBC posteriors might not be identical to those from data with no glitches. This is unsurprisingly most prominent for signals and glitches with similar time-frequency morphology. Despite this, the CBC parameters are correctly estimated within the extent of the posterior. As a point of comparison for each case, we also examine the bias incurred on CBC parameters by performing a standard analysis that ignores the presence of the glitch in the data entirely. 

Our process allows us to obtain both a model for the glitch that can be subtracted from the data and parameter estimation results for the CBC signal, though the latter are restricted by the assumption of aligned-spins in the current algorithm implementation. The glitch-subtracted data are ready for downstream analyses with more sophisticated waveform or detector calibration models. We demonstrate that the lack of spin-precession and higher-order modes in our CBC waveform models does not hinder accurate CBC-glitch separation. We also test whether we can do glitch subtraction on single-detector data. The additional examples and checks presented here demonstrate that the analysis is ready to tackle incoming data in O4~\cite{Aasi:2013wya}.

Finally, we consider some low-significance triggers and attempt to distinguish between CBC signals, Gaussian noise, glitches, or a combination of all three present in the data. Standard detection efforts consider only the possibility that either a CBC signal or a glitch is present in some data. It is therefore possible that the significance of a CBC signal could be impacted if it overlaps with a glitch as this could make the data inconsistent with our CBC model. We find that trigger 191225\_215715 (hereon S191225)~\cite{gwtc3} and trigger 200114\_020818 (hereon  S200114)~\cite{LIGOScientific:2021tfm} are consistent with the presence of both a glitch and a CBC signal. The above results are subject to the caveat that we use a CBC waveform model without spin-precession or higher-order modes.

The paper continues as follows. In Sec.~\ref{sec:alg}, we describe the \BayesWave algorithm, focusing on the CBC (templated) and the glitch models. In Sec.~\ref{sec:method} we present our injection and recovery scheme. In Sec.~\ref{sec:glitches} we present our results on an array of different glitch classes. We test the systematics and limitations of our CBC sampler in Sec.~\ref{sec:futherStudies} by including injections containing higher order modes, spin-precession, and events in a single detector. Finally, in Sec.~\ref{sec:signalClassification}, we apply our analysis on triggers S191225 and S200114 in order to assess the presence of signals and/or glitches in the data. In Sec.~\ref{sec:conclusions} we conclude.


\section{General Methodology Description}\label{sec:alg}
In this section we describe the main features of our analysis that estimates parameters of GW events using templates while jointly modeling detector glitches using wavelets. We briefly introduce our inference scheme, then discuss the features of {\tt BayesWave} relevant for this study including the models, priors, and the joint sampler.

\subsection{Brief Introduction to Inference Scheme}
\label{subsec:inference}

GW parameter estimation aims to compute $p(\theta_h|d)$, the posterior probability distribution that a model $h$ with parameters $\theta_h$ describes the given data, $d$. The quantity $h$ can contain any component of the data we attempt to model, for example a CBC signal, a glitch, or Gaussian noise. In the case of CBC signals, $h$ is typically a waveform template and $\theta_h$ are parameters that describe the binary. The posterior $p(\theta_h|d)$ is proportional to the prior $p(\theta_h)$ times the likelihood of observing data $d$ given the model $h(\theta_h)$, $p(d|\theta_h)$. The likelihood function encodes our assumption about the detector noise. For LIGO analyses and stationary, Gaussian noise with power spectral density (PSD) $S_n(f)$ in each detector, the likelihood is given by Eq. 3 in~\cite{Cornish:2020dwh}. Under this formulation, the multidimensional parameter posterior $p(\theta_h|d)$ is typically explored with stochastic sampling methods such as Markov Chain Monte Carlo (MCMC) and nested sampling~\cite{Veitch:2014wba,Romero-Shaw:2020owr}.

\begin{figure*}[ht]
	\centering
	\includegraphics[]{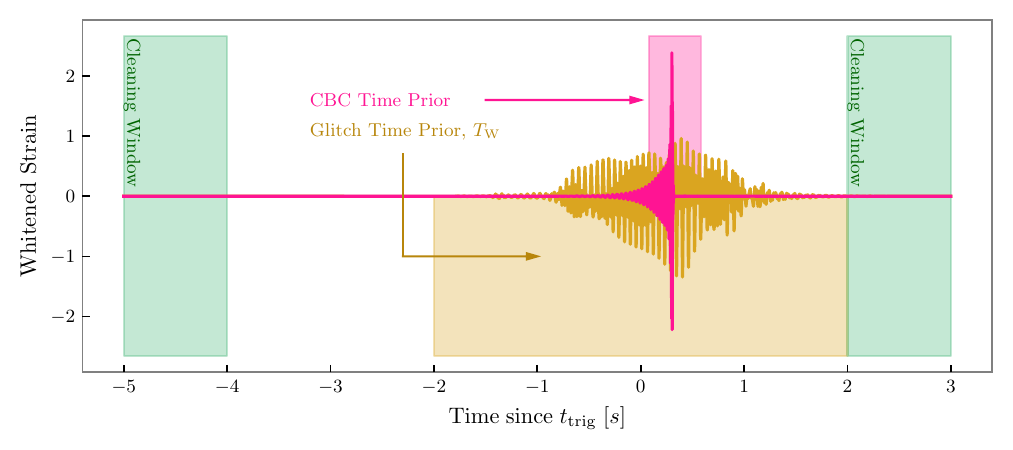}
	\caption{Analysis setup with the different time intervals and priors for the analysis of a CBC signal (magenta) and a glitch (gold). 
		The CBC (glitch) time prior is depicted by the magenta (gold) shaded region, although the two regions need not share a common center. The preprocessing cleaning phase is used to permanently subtract wavelets from the data in the green region.}
	\label{fig:windows}
\end{figure*}

\subsection{{\tt BayesWave} Analysis}

\BayesWave~\cite{Cornish:2014kda,Littenberg:2014oda,Cornish:2020dwh,bayeswave} is a flexible data analysis algorithm that models various components in GW data including signals, glitches, and noise. \BayesWave explores the joint posterior distribution of its models using a combination of MCMC and reversible jump (RJ) MCMC~\cite{10.1093/biomet/82.4.711} samplers. The algorithm has a wide range of applications including an unmodeled search pipeline~\cite{LIGOScientific:2019ppi, LIGOScientific:2021hoh}, on-source PSD generation~\cite{O3a_catalogue, gwtc3,Chatziioannou:2019}, tests of General Relativity~\cite{TheLIGOScientific:2016src, LIGOScientific:2019fpa, LIGOScientific:2020tif,Chatziioannou:2021mij} and consistency tests~\cite{Ghonge:2020suv}, and various studies of poorly modeled or unmodeled sources including binary neutron star (BNS) postmergers~\cite{Chatziioannou:2017ixj, Abbott:2018wiz, Torres-Rivas:2018svp}, eccentric BBH mergers~\cite{dalya:2020}, and supernova~\cite{Raza:2022kcs}. \BayesWave has also been used to perform glitch subtraction~\cite{TheLIGOScientific:2017qsa,Pankow:2018qpo,O3a_catalogue, gwtc3}. With this flexibility to model a wide range of signal and glitch morphologies in hand, in Ref.~\cite{Chatziioannou:2021ezd} we extended the algorithm capabilities to include a model for CBC signals based on CBC waveform templates. The algorithm implementation is described in detail in~\cite{Cornish:2014kda,Littenberg:2014oda,Cornish:2020dwh,Wijngaarden:2022sah} and we briefly summarize the most relevant features here.
 
Whereas most GW parameter estimation analyses~\cite{Veitch:2014wba,Romero-Shaw:2020owr} use single, point estimates of the PSD and assume that the data contains no glitches, \BayesWave can relax both assumptions by modeling the CBC signal, the noise PSD, and any potential transient noise all at once.\footnote{The effect of PSD uncertainty in CBC analyses has also been explored in~\cite{2013PhRvD.88h4044L, Talbot:2020auc, Biscoveanu:2020kat, Edwards:2015eka, Banagiri:2019lon, Ashton:2020mrh, Vajpeyi:2021ccn, Rover:2008yp, Rover:2011qd}.} The full \BayesWave model for this study is the union of a CBC (waveform template) model, a glitch model, and a noise PSD model. 
The different \BayesWave models are described below. 

\begin{itemize}
	\item The \textit{noise model}, or PSD model, expresses the noise PSD as a sum of splines and Lorentzians. The splines capture the smooth underlying broadband noise whereas the Lorentzians capture the sharp spectral peaks. Within this paper, we use the color grey to represent the noise model. 
	\item The \textit{glitch model} expresses excess detector transient noise as the sum of sine-Gaussian wavelets: accordingly, each detector has its own, independent, set of glitch wavelets. The set of all sine-Gaussian wavelets form an overcomplete basis over smooth function space, and are thus able to describe a glitch of any morphology provided that it is sufficiently loud. The number of wavelets and hence the model dimensionality is not fixed and wavelets can be added or subtracted as needed. Each wavelet is described by five parameters ($\theta^{\text{glitch}}:$) central time, frequency, quality factor, amplitude, and phase. The quality factor is related to the damping time of the sine-Gaussian, and together with the frequency determines the duration of each wavelet. The functional form of the wavelet is given by Eq. 4 in~\cite{Cornish:2020dwh}. Within this paper, we use gold to represent the glitch model. 
	\item The \textit{CBC model} uses waveform templates to capture the CBC signal in a manner similar to traditional GW parameter estimation~\cite{Veitch:2014wba,Romero-Shaw:2020owr}. Details of the implementation of the CBC model in \BayesWave are given in~\cite{Wijngaarden:2022sah}. For this analysis we restrict to quasicircular CBC signals whose spins are aligned with the orbital angular momentum. Such signals are characterized by up to 13 parameters, namely four intrinsic CBC parameters (the two masses $m_1, m_2$ and two dimensionless spins $\chi_1,\chi_2$ projected onto the Newtonian orbital angular momentum) and seven extrinsic parameters (a time and phase, the right ascension and declination, the luminosity distance $D_L$, the inclination $\iota$, and the polarization angle). For binary neutron stars (BNSs), we also have two tidal parameters $\Lambda_1, \Lambda_2$. In what follows, we express the masses through the chirp mass ${\cal{M}}_c \equiv \left(m_1 m_2\right)^{3/5}\left(m_1 + m_2\right)^{-1/5}$ which determines the GW phase evolution to leading order and the mass ratio $q\equiv m_2/m_2<1$. We also express the spin through $\chieff\equiv (m_1\chi_1+m_2\chi_2)/(m_1+m_2)$, which is conserved approximately throughout the binary inspiral~\cite{Racine:2008qv}. Within this paper, we use pink to represent the CBC model. 

	\item Though not used here, for completeness we also mention the \textit{signal model} that fits for coherent, excess power (``unmodeled" astrophysical signals) using again sine-Gaussian wavelets. Unlike the glitch model, the signal model enforces that the wavelets must be coherent across the detector network as a genuine astrophysical signal would be. Both the signal and the CBC model have the potential to capture a CBC source though the former is more flexible, and thus less sensitive, particularly to weak or long-duration signals.
\end{itemize}
%

\subsection{Priors}
\label{subsec:priors}

The priors for the glitch and CBC model parameters remain mostly unchanged compared to~\cite{Chatziioannou:2021ezd,Wijngaarden:2022sah}. However, for some combinations of glitch and CBC signals, we find that additional flexibility is required in the time placement of glitch wavelets and the CBC template. By construction, \BayesWave analyzes data of duration $\Tobs$ around some trigger time $\ttrig$. The prior on the central time of the glitch wavelets then has support within a ``window" of length $\Tw<\Tobs$ around $\ttrig$, while the CBC time prior is by default $(\ttrig-0.5\mathrm{s},\ttrig+1.5\mathrm{s})$. Here we relax the requirement that glitches and CBC signals have a time prior around a common time $\ttrig$ and allow for them to be placed in different time intervals, though still within $\Tobs$. The priors for the noise PSD model are all unchanged compared to~\cite{Littenberg:2014oda,Chatziioannou:2021ezd}. Figure~\ref{fig:windows} shows the relation between the different time intervals.

Though the wavelets that model the glitch can have central times only within $\Tw$, the current \BayesWave implementation employs a preprocessing ``cleaning phase". During this phase, the algorithm is run with only the glitch model activated and $\Tw=\Tobs$, i.e., wavelets can be placed anywhere in the analyzed data segment. At the end of the cleaning phase and before proceeding to the main analysis, wavelets with time centered within a specified interval are permanently subtracted from the data. We typically use a $1$s cleaning window at the beginning and end of the data segment. This procedure removes any glitch power that might be present in the analysis segment but not necessarily close to the CBC signal itself as well as data artifacts caused by the finite segment duration and that could bias the PSD estimation.

\subsection{Sampler}
\label{subsec:sampler}

\begin{figure}[h]
	\includegraphics[width=0.45\textwidth]{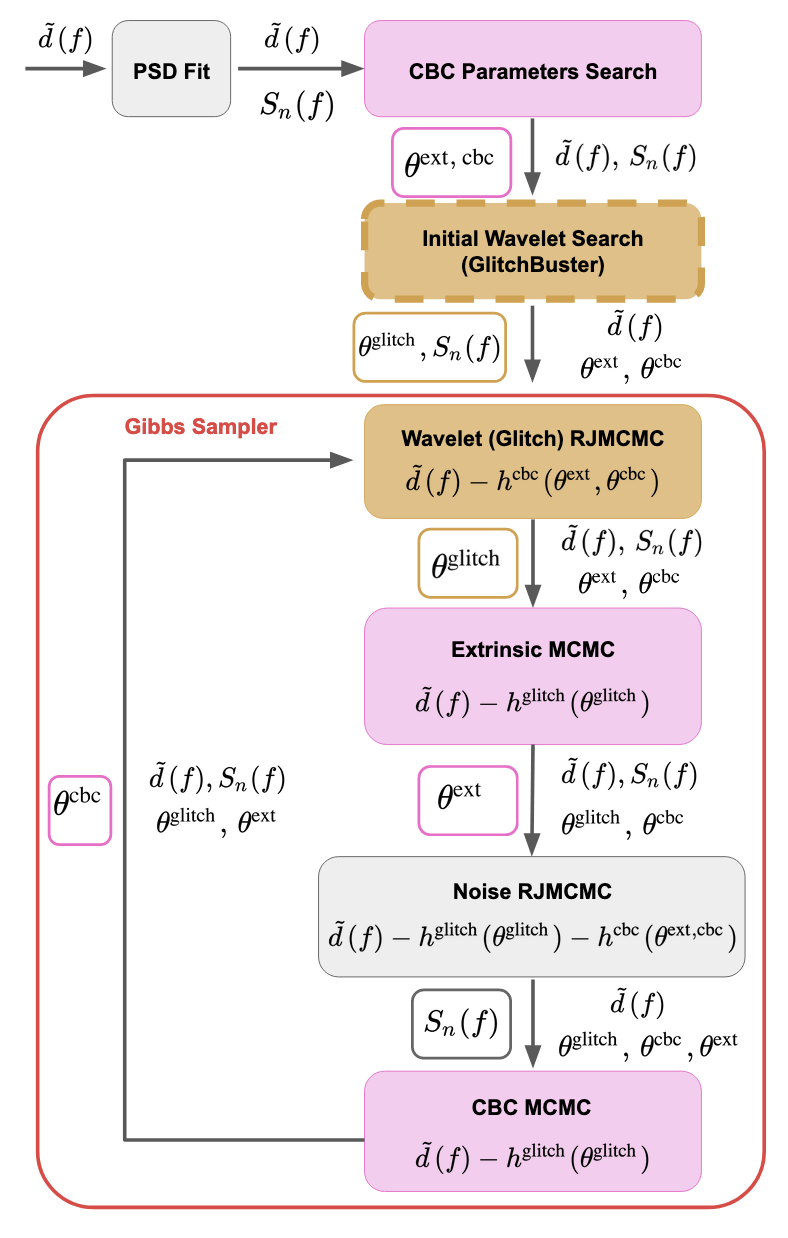}
	\caption{ Visual depiction of the BayesWave workflow described in Sec. \ref{sec:alg}. Each colored box represents a component sampler and displays its name (bolded) and its input data (in LaTeX). Pink (gold, grey) boxes indicate the sampler that searches over the CBC (glitch, PSD) parameter space. Underneath each colored box, the sampled (fixed) parameters are boxed (unboxed). The Gibbs sampler (i.e., the cycle of component samplers) is boxed in red. Before entering the Gibbs sampler, we use the Fourier domain strain data, $\tilde{d}(f)$, to obtain initial points for the noise and CBC samplers through the procedure described in~\cite{Cornish:2021wxy}. We also optionally (dotted box) obtain an initial point for the glitch model~\cite{Cornish:2020dwh}, otherwise the glitch model begins with 0 wavelets. Within the Gibbs sampler, each component (RJ)MCMC goes through $\mathcal{O}(10^2)$ iterations before passing its values to the next sampler. After each Gibbs sampler loop, a single sample for all parameters is returned. The Gibbs sampler loops $\mathcal{O}(10^4)$ times.}\label{fig:workflow}
\end{figure}

To characterize the joint glitch, CBC, and noise posterior, \BayesWave uses a combination of MCMC and RJMCMC samplers stringed together within a blocked Gibbs sampler. The blocked samplers give us the flexibility to trivially turn models on and off during an analysis.
Here we sketch the workflow for the ``CBC+Glitch" analysis, but other \BayesWave running modes with different model combinations vary only in which samplers are active.

Instead of sampling all parameters concurrently, we separate them into groups of related parameters that are sampled together while other parameters are held fixed. The order of the corresponding samplers and a breakdown of the fixed or varying parameters within each sampler is displayed in Fig.~\ref{fig:workflow}. Each component sampler runs for a predetermined number of iterations, typically $\mathcal{O}(100)$ and returns its last iteration to be used as fixed parameters by the other samplers. We iterate through the Gibbs sampler loop $\mathcal{O}(10^4)$ times before completing. 

Before the Gibbs sampler begins, each model (i.e. PSD, CBC, and glitch) needs to be initialized. An initial estimate for the PSD is generated by the methods described in~\cite{Cornish:2020dwh}. To initialize the CBC parameters we follow~\cite{Cornish:2021wxy}. An optional \texttt{GlitchBuster} step finds initial parameters for the glitch wavelets by iteratively estimating the PSD, wavelet-transforming the data, and removing excess power wavelets~\cite{Torrence:1998}. The procedure is described in more detail in~\cite{Cornish:2020dwh}. Without \texttt{GlitchBuster}, the glitch model begins with no wavelets. 

The sampling procedure, and specifically the integration of the CBC sampler is described in detail in~\cite{Wijngaarden:2022sah}. Below we briefly describe each individual sampler. 

\begin{itemize}
	\item The ``wavelet (glitch) RJMCMC" updates the parameters of one of the current wavelets or adds/subtracts a wavelet. Details of the RJMCMC implementation are presented in~\cite{Cornish:2014kda,Cornish:2020dwh}.
	\item The ``extrinsic MCMC" updates the extrinsic parameters of the signal, namely the distance, sky-location, inclination and polarization angles, and time\footnote{Though not used in the study, the \BayesWave \emph{signal model} that describes a GW signal with coherent sine-Gaussian wavelets also makes use of the extrinsic sampler.}. Details are provided in~\cite{Cornish:2014kda,Cornish:2020dwh, Wijngaarden:2022sah}.
	\item The ``CBC MCMC" updates the intrinsic CBC parameters (masses, spins, tides) as well as parameters that are correlated with them, namely the distance, time, and phase. We use waveforms implemented in the {\tt LALSimulation} suite of models~\cite{lalsuite}. The sampler can operate with any non-precessing model available in {\tt LALSimulation} and in this study we rely on {\tt IMRPhenomD}~\cite{Husa:2015iqa,Khan:2015jqa}. We currently do not account for nonaligned spin degrees of freedom but plan to extend our analysis to include the effect of spin-precession in the future. The sampling proposals as well as the heterodyne procedure~\cite{Cornish:2021wxy,Cornish:2021lje} used to speed up the likelihood calculation are described in~\cite{Wijngaarden:2022sah}.
	\item The ``noise RJMCMC" updates the number and parameters of the splines and Lorentzians that describe the noise PSD. Details are provided in~\cite{Littenberg:2014oda}. 
\end{itemize}


\section{Data and Analysis Setup}
\label{sec:method}

\begin{figure*}
	 \includegraphics[]{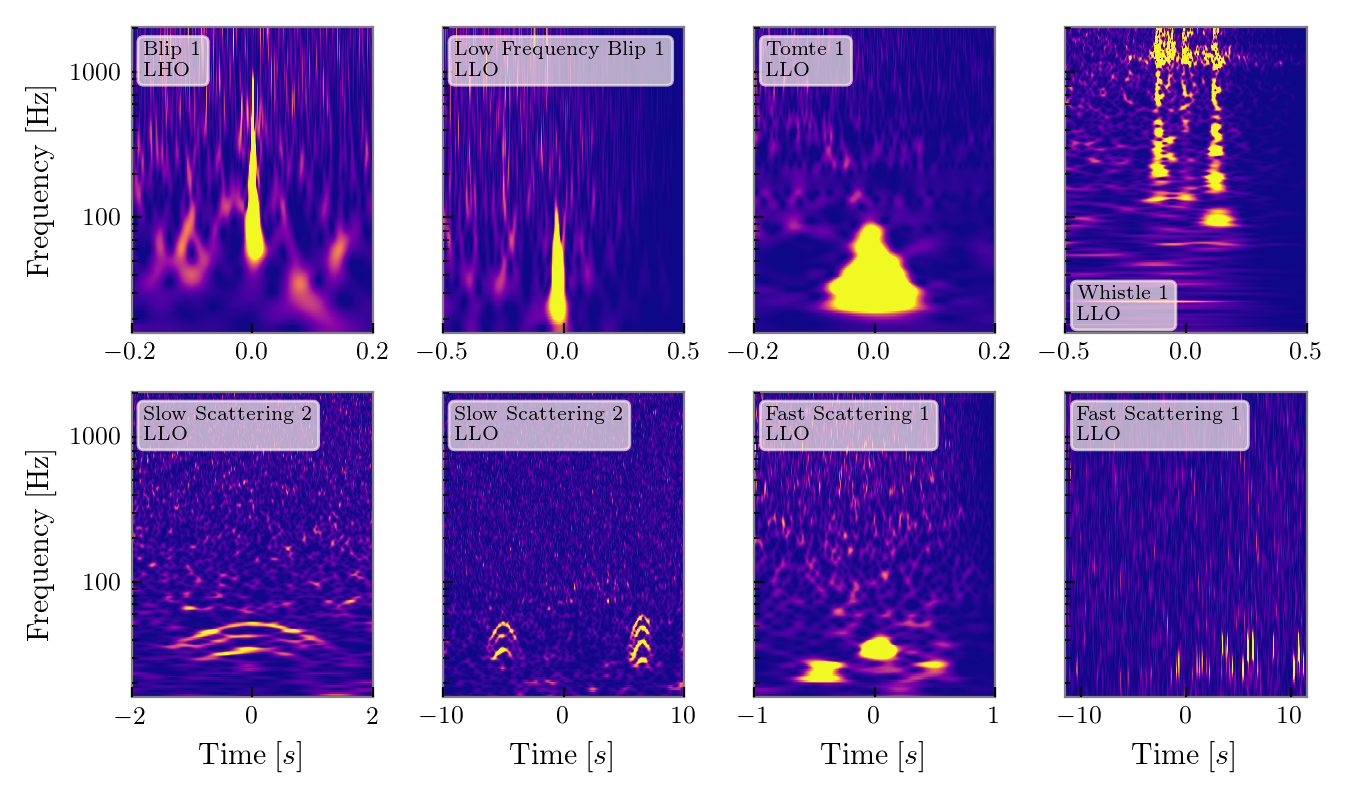}
	\caption{Spectrograms of representative glitches for each glitch family we consider in our study. We display two spectrograms for the same fast and slow scattering glitches to demonstrate the long- and short-term behavior of such glitch types. See Tab.~\ref{tab:settings} for glitch GPS times.}\label{fig:Spectrograms_ALL}
\end{figure*}

\begin{table}
	\begin{tabular}{c|ccccccc}
		Signal & $m_1 (M_\odot)$ & $m_2 (M_\odot)$ & min($\Tobs$) (s) & $\Lambda_1$ & 
		$\Lambda_2$ \\
		\hline
		HM BBH  & 36  & 29       & 4      & -  & -     \\
		LM BBH   & 12  & 7        & 16     & -  & -     \\
		BNS            & 1.5 & 1.4      & 128    & 115 & 320               
	\end{tabular}
	\caption{Parameters of the injected signals we consider. For all the injections we set $\chieff = 0$, $\cos{\iota} = 0.88$, $\phi_0 = 1.23$, and $\psi = 0.3$. The sky location is fixed overhead LLO, while the distance is varied to keep the network SNR fixed at $\sim15$. For each injection type, we display the minimum segment length $\Tobs$ necessary to contain the signal, but certain long-duration glitches required an increased segment length.} \label{tab:injtable}
\end{table}

\begin{table}
	\begin{tabular}{c|c|c|c}
		Quantity &  Data & Recovery & Description \\
		\hline
		 $\gCBC$ & glitch and  CBC & CBC+Glitch & recovered glitch \\
		 $\gG$ & glitch  & GlitchOnly & recovered glitch \\   
		 $\cbcG$  & glitch and  CBC & CBC+Glitch & recovered CBC \\
		 $\cbcCBC$ & glitch and CBC & CBCOnly & recovered CBC  \\  
		 $\inj$ & N/A & N/A & injected CBC
	\end{tabular}
	\caption{Summary of quantities used in the overlaps. From left to right columns provide the symbol, the relevant data, the models active during recovery, and a description of what each quantity is.} \label{tab:overlaps}
\end{table}

To test the efficacy of our analysis, we use LIGO data from O2 and O3, available through the GW Open Science Center~\cite{LIGOScientific:2019lzm}. 
Since we do not have exact first-principles models for glitches\footnote{Phenomonological models for some glitch types have been constructed, for example~\cite{Merritt:2021xwh}.}, we identify data containing genuine detector glitches, which were classified through Gravity Spy~\cite{GravitySpy}.
We then inject a known CBC signal on top of the glitch, and analyze the data by simultaneously modeling the CBC, glitch, and noise. We label such analyses as ``CBC+Glitch". In each case we also analyze the same data using only the CBC and noise models, i.e., ignoring the presence of the glitch. We label such analyses as ``CBCOnly".  To create a point of reference for the glitch, we also consider the original data with no CBC injection with a ``GlitchOnly" run using only the glitch and noise models. 

Going beyond the study of~\cite{Chatziioannou:2021ezd}, here we analyze a more extensive set of glitch types (as classified by GravitySpy \cite{Soni:2021cjy, Zevin:2016qwy}), glitch instances, and a set of both BBH and BNS injections. The injected signals include high-mass (HM) BBH (GW150914-like~\cite{Abbott:2016blz}), low-mass (LM) BBH (GW170608-like~\cite{Abbott:2017gyy}), and BNS (GW170817-like~\cite{TheLIGOScientific:2017qsa}) injections; their parameters are provided in Table~\ref{tab:injtable}. Though initially we targeted specific glitches, in many cases secondary glitches (occurring in either or both detectors) were present in the data, speaking to the high occurrence rate of glitches. In those cases we do not discard the data; we analyze them nonetheless and attempt to model all glitches. These analyses also show that \BayesWave can differentiate between signals and glitches even when they occur in multiple detectors.
A spectrogram of one glitch per glitch family is shown in Fig.~\ref{fig:Spectrograms_ALL}.
Although the ``worst case scenario" for CBC analyses is a glitch with SNR similar to or greater than that of the signal that has a similar time-frequency morphology, in our validation studies we include a wide range of combinations of signals and glitches.

We fix the intrinsic parameters and the sky location within injections of the same CBC type. For each combination of glitch and CBC type, we inject the signal at different times with respect to the glitch in order to vary the amount of overlap between the two. Each signal has a network signal-to-noise ratio (SNR) of 15 and we use the {\tt IMRPhenomD}~\cite{Husa:2015iqa,Khan:2015jqa} ({\tt IMRPhenomD\_NRTidalv2}~\cite{Dietrich:2017aum,Dietrich:2018uni,Dietrich:2019kaq}) model for the BBH (BNS) signals for injection and recovery unless otherwise indicated. For computational efficiency and since the BNSs overlap with all glitches at low frequencies (below $40$Hz), we use a low sampling rate that does not allow us to extract tidal parameters. The possibility of separating BNSs and glitches when they overlap in the high frequencies relevant for tidal effects was shown in~\cite{Chatziioannou:2021tdi} using the same analysis as here.

The recovered CBC parameters can be compared with the injected ones directly to assess the recovery reliability. However, no such comparison is possible for glitches as they are obtained from real data. We therefore use various overlaps (${\cal{O}}$, defined in Eq. 8 of~\cite{Ghonge:2020suv}) and mismatches (one minus the overlap, ${\cal{M}}=1-{\cal{O}}$) between the recovered CBC and glitch to quantify the quality of the CBC-glitch separation. Definitions are given in Table~\ref{tab:overlaps}. Specifically, for each glitch we also analyze the original data with no CBC injection using only the glitch and noise models. This gives us access to a baseline estimate for the glitch that can be compared to estimates of the same glitch recovered in data coincident with a signal. We then compute the mismatch between the median glitch recovery from analyses with ($\gCBC$) and without ($\gG$) injected CBC signals as a way to test whether the recovered glitch subsumes part of the CBC signal power. A high mismatch could suggest that either the glitch model captures part of the CBC (and leads to it being inadvertently subtracted from the data) or the glitch model misses part of the glitch and fails to subtract all glitch power.
Additionally, we compute the posterior for ${\cal{M}}(\cbcCBC|\inj)$, the mismatch between the injected and the recovered CBC signal whose expected value is a function of SNR~\cite{Chatziioannou:2017tdw}.

\begin{figure}
	\includegraphics[width=\linewidth]{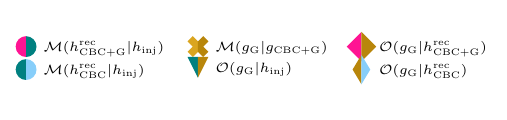}
	\includegraphics[width=\linewidth]{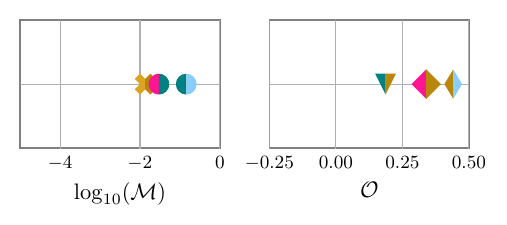}
	\caption{Example result for the mismatches defined in Table~\ref{tab:matches}. In this case the glitch recovery is not impacted by the presence of the CBC as $\MgGgCBC$ (gold cross; left panel) is low. Additionally $\McbcGinj<\McbcCBCinj$ (magenta/teal and blue/teal; left panel) so the CBC model is recovering part of the glitch if one uses an analysis that assumes pure Gaussian noise. We see $\OgGinj<\OgCBCgG<\OcbcCBCgG$ (teal/gold, magenta/gold, and blue/gold; right panel) meaning that the CBC model is absorbing part of the glitch power, but doing less so when the glitch is part of the model.  
	However, such overlaps are quite low, so the effect is small. }
	\label{fig:overlaps_example}
\end{figure}

In each case we also carry out a ``CBCOnly" analysis that is akin to traditional parameter estimation (with the difference that we also marginalize over the PSD). Though such an analysis is not well motivated as the model cannot exactly match the data (and we encounter increased convergence issues), we find it instructive to compare its results to the full ``CBC+Glitch" analysis. We do so by plotting both posteriors when discussing each glitch, and also via the Jensen-Shannon divergence in Sec.~\ref{sec:JSDivergence}. We denote the recovered CBC signal from such an analysis as $\cbcCBC$ and compute ${\cal{M}}(\gG|\cbcCBC)$ as an estimate of how much of the glitch the CBC model can recover. We also compare ${\cal{M}}(\cbcG|\inj)$ against  ${\cal{M}}(\cbcCBC|\inj)$  to test the effect of assuming pure Gaussian noise on CBC recovery in the presence of a glitch. An example result for the various overlaps is presented in Fig.~\ref{fig:overlaps_example} and described in detail in Table~\ref{tab:matches}. Generally speaking, good CBC-glitch separation is achieved when the quantities on each panel of Fig.~\ref{fig:overlaps_example} and equivalent figures in later sections are small.

%

\newcommand\cincludegraphics[2][]{\raisebox{-0.5\height}{\includegraphics[#1]{#2}}}
\begin{table*}
	\begin{tabular}{p{0.20\linewidth} | p{0.06\linewidth} | p{0.7\linewidth}}
		\hline
		Mismatch   & Symbol & Interpretation \\
		\hline
		$\MgGgCBC$  & \cincludegraphics[width=9mm, height=9mm]{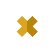} & Mismatch between the median glitch reconstruction from data with and without a CBC injection. A high value means that the recovered glitch has been impacted by the presence of the CBC. \\
		$\McbcGinj$  & \cincludegraphics[width=9mm, height=9mm]{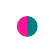} & Median mismatch between injected and recovered CBC signal when accounting for the glitch. The lower its value, the better the CBC recovery, though the expected value depends on the signal SNR. \\
		$\McbcCBCinj$  & \cincludegraphics[width=9mm, height=9mm]{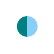} & Median mismatch between injected and recovered CBC signal without accounting for the glitch. If $\McbcGinj<\McbcCBCinj$, then the CBC model is subsuming glitch power when the latter is left unaccounted for.\\
		$\OgGinj$  & \cincludegraphics[width=9mm, height=9mm]{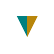} & Overlap between injected CBC signal and the median glitch recovered without a CBC injection. The absolute value of $\OgGinj$ indicates how similar the CBC and the glitch are, and thus how difficult the separation will be. \\
		$\OgCBCgG$  & \cincludegraphics[width=9mm, height=9mm]{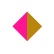} & Overlap between the median CBC recovered when accounting for glitch power and the median glitch recovered without a CBC injection. If $\OgCBCgG>\OgGinj$, then the CBC model might be absorbing undue glitch power.  \\
		$\OcbcCBCgG$  & \cincludegraphics[width=9mm, height=9mm]{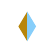} & Overlap between the median CBC recovered without accounting for glitch power and the median glitch recovered without a CBC injection. If $\OcbcCBCgG > \OgGinj$, then the CBC model might be absorbing undue glitch power. \\
		\hline
	\end{tabular}\caption{Detailed description of the various mismatches we compute using the reconstructions defined in Table~\ref{tab:overlaps}. An example result for these mismatches is plotted in Fig.~\ref{fig:overlaps_example}. Throughout, pink and light blue are used for CBC reconstructions with and without the glitch model respectively, gold is used for glitch reconstructions, and teal is used for CBC injections. The split colors indicate the two models used for the overlap or mismatch.}\label{tab:matches}
\end{table*}

Another potential test that was explored over the course of this analysis is the Anderson-Darling statistic~\cite{AndersonDarling}. This test can be used to assess the degree of Gaussianity in the residual and has been proposed in the context of PSDs~\cite{Chatziioannou:2019}. Specifically, we explored the option of subtracting some point estimate (such as the median, or a fair posterior draw) for the CBC and the glitch models from the data and then computing the Anderson-Darling statistic. However, we found that the test is very forgiving and even fails to identify large amounts of residual glitch power. We attribute this to the fact that the form of the Anderson-Darling test we employ is better suited for identifying large non-Gaussian tails in distributions (and thus it is well suited for PSDs~\cite{Chatziioannou:2019, Talbot:2020auc}) than for coherent non-Gaussian residual that affects only a few data points.

\begin{table*}
	\begin{tabular}{c|cccccccc}
		\hline
		Run Label   & GPS Time & Injected signal   & Glitches           &   $\Tobs$[s] &   $\Tw$[s] &   $\Qmax$ &   $\Dmax$ &   $\flow$[Hz] \\
		\hline
		{\tt B1\_HM\_BBH}  & 1168989748 & HM BBH   & Blip LHO         &      4  &  1   &     40 &    100 &     16 \\
		{\tt B1\_LM\_BBH}   & 1168989748 & LM BBH   & Blip LHO         &     16  &  1   &     40 &    100 &     16 \\
		{\tt B1\_BNS}           & 1168989748 & BNS          & Blip LLO          &    128 &  1   &     60 &    100 &     16 \\
		\hline
		{\tt B2\_HM\_BBH}   & 1165578732 & HM BBH   & 2$\times$Blip LHO         &      4 &  1   &     40 &    100 &     16 \\
		{\tt B2\_LM\_BBH}   & 1165578732 & LM BBH   & 2$\times$Blip LHO          &     16 &  1   &     40 &    100 &     16 \\
		\hline
		{\tt B3\_HM\_BBH}   & 1171588982 & HM BBH   & Blip LLO / Unclassified LHO    &      4 &  1   &     40 &    100 &     16 \\
		{\tt B3\_LM\_BBH}   & 1171588982 & LM BBH   & Blip LLO  / Unclassified LHO         &     16 &  1   &     40 &    100 &     16 \\
		\hline
		{\tt S1\_HM\_BBH}   & 1172917780 & HM BBH   & Slow Scattering LLO     &      8 &  4   &    160 &    100 &      8 \\
		{\tt S1\_LM\_BBH}   & 1172917780 & LM BBH   & Slow Scattering LLO    &     16 &  4   &    160 &    100 &       8 \\
		\hline
		{\tt S2\_HM\_BBH}   & 1166358283 & HM BBH   & Slow Scattering / Blip / High Frequency Lines LLO  &      8 &  4   &    250 &    100 &     16 \\
		{\tt S2\_LM\_BBH}   & 1166358283 & LM BBH   & Slow Scattering / Blip / High Frequency Lines LLO     &     16 &  4   &    250 &    100 &     16 \\
		{\tt S2\_BNS}      & 1166358288 & BNS      & 2$\times$Slow Scattering / Blip / High Frequency Lines LLO   &    128 & 30   &    250 &    100 &     16 \\
		\hline
		{\tt S3\_HM\_BBH}   & 1177523957 & HM BBH   & Slow Scattering LLO     &      8 &  4.5 &    250 &    100 &     8 \\
		{\tt S3\_LM\_BBH}   & 1177523957 & LM BBH   & Slow Scattering LLO     &     16 &  4.5 &    250 &    100 &     8 \\
		\hline
		{\tt FS1\_HM\_BBH}  & 1238326223 & HM BBH   & Fast Scattering LLO &      4 &  2   &     60 &    100 &     16 \\
		{\tt FS1\_LM\_BBH}  & 1238326212 & LM BBH   & Fast Scattering LLO   &     32 & 23   &     60 &    200 &     16 \\
		{\tt FS1\_BNS}      & 1238326221 & BNS   & Fast Scattering LLO              & 128     &  84.7   &    60 &    100 &     16 \\
		\hline
		{\tt FS2\_HM\_BBH}  & 1265656683 & HM BBH   & Fast Scattering LLO &     4 &  3     &     60 &    100 &     16 \\
		{\tt FS2\_LM\_BBH}  & 1265656673 & LM BBH   & Fast Scattering LLO   &  64 &  32   &    60 &    100 &     16 \\
		\hline
		{\tt FS3\_HM\_BBH}  & 1266384078 & HM BBH   & Fast Scattering LLO/Blip LHO &      4 &  3   &     60 &    100 &     16 \\
		{\tt FS3\_LM\_BBH}  & 1266384070 & LM BBH   & Fast Scattering LLO/Blip LHO &     32 & 30   &     60 &    100 &     16 \\
		\hline
		{\tt BL1\_HM\_BBH}  & 1256909978 & HM BBH   & Low-Frequency Blip LLO    &      4 &  1   &     40 &    100 &     16 \\
		\hline
		{\tt T1\_HM\_BBH}   & 1243679046 & HM BBH   & Tomte LLO/Blip LHO        &      4 &  1   &     40 &    100 &      8 \\
		{\tt T1\_LM\_BBH}   & 1243679046 & LM BBH   & Tomte LLO/Blip LHO        &     16 &  3   &    100 &    100 &     16 \\
		\hline
		{\tt W1\_HM\_BBH}   & 1253426470 & HM BBH   & Whistle LLO        &      4 &  1   &    100 &    400 &     16 \\
	\end{tabular}
	\caption{
		Settings for the analyses of Sec.~\ref{sec:glitches}. From left to right, columns correspond to the run label used in subsequent plots, the approximate GPS time of the given glitch (center of $\Tw$ as in Fig.~\ref{fig:windows}), the type of injected signal, the glitches encountered and the affected detector(s), the segment length $\Tobs$, the duration of the time window where glitch wavelets can be placed $\Tw$, the maximum quality factor of the glitch wavelets $\Qmax$, the maximum number of wavelets allowed $\Dmax$, and the low-frequency cutoff $\flow$. All BBH runs have a sampling rate of 2048Hz whereas the BNS runs used a sampling rate of $1024$ Hz (which precludes any possibility of recovering tidal parameters). Horizontal lines group analyses that target overlapping data, though the exact center of the glitch window (GPS time) might be shifted according to long- and short-term glitch behavior.
	}
	\label{tab:settings}
\end{table*}


\section{Results on overlapping signals and glitches}
\label{sec:glitches}

In this section we present results from our injection study. Each subsection corresponds to a different glitch family and injections using three different CBC sources, see Table \ref{tab:injtable}. In many cases, especially for the longer signal analyses, the data contain additional glitches beyond what was intended. We point these cases out, but our inability to easily find glitch-free data of duration $1-2$ minutes speaks to their prevalence in LIGO data. Details about the times analyzed and the glitches that we encountered either intentionally or accidentally are provided in Table~\ref{tab:settings}, together with analysis settings. 

\subsection{Blip Glitches}\label{sec:Blip}

\begin{figure*}
	\includegraphics[width=0.85\linewidth]{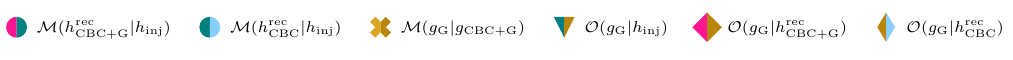}
	\includegraphics[width=0.85\linewidth]{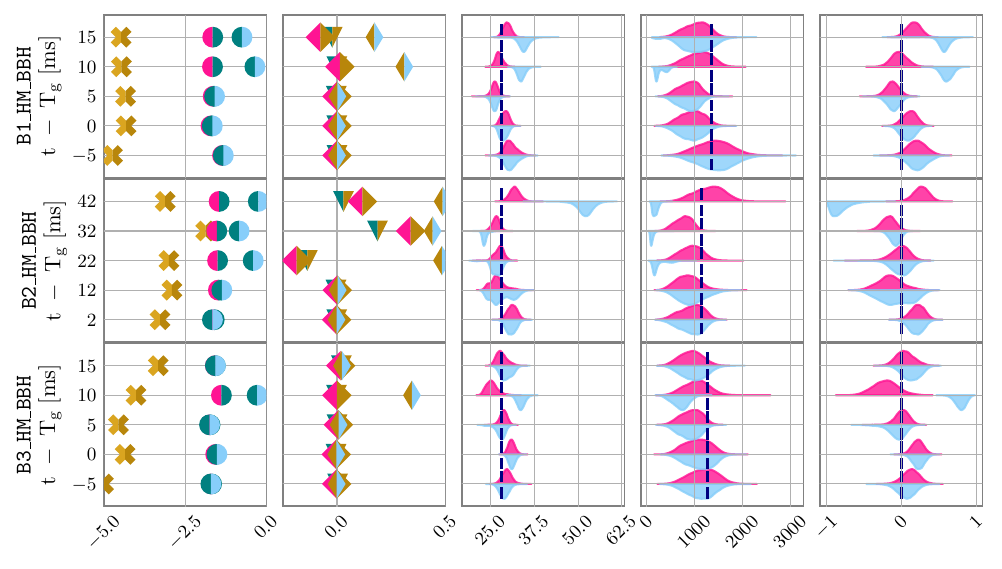}
	\includegraphics[width=0.85\linewidth]{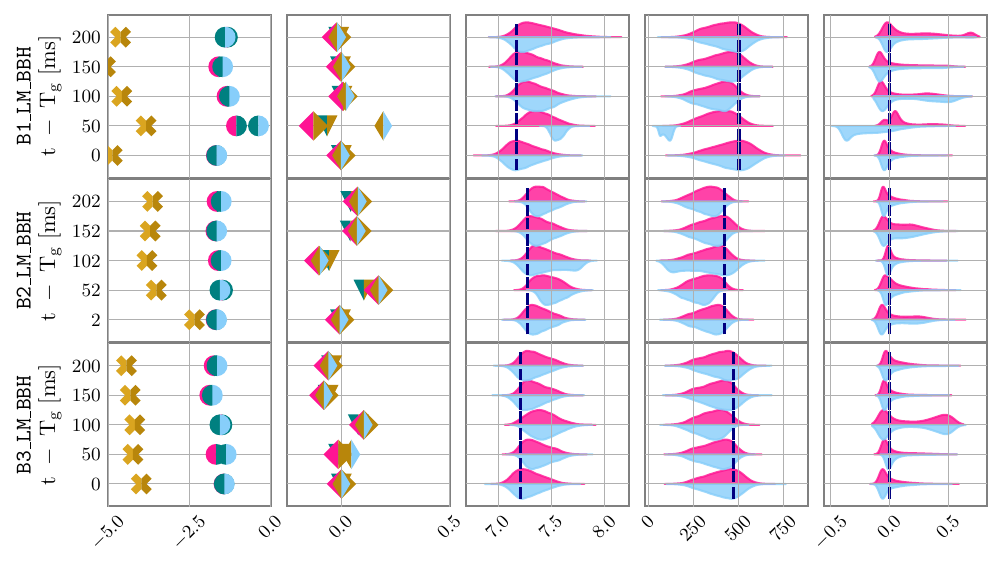}
	\includegraphics[width=0.85\linewidth]{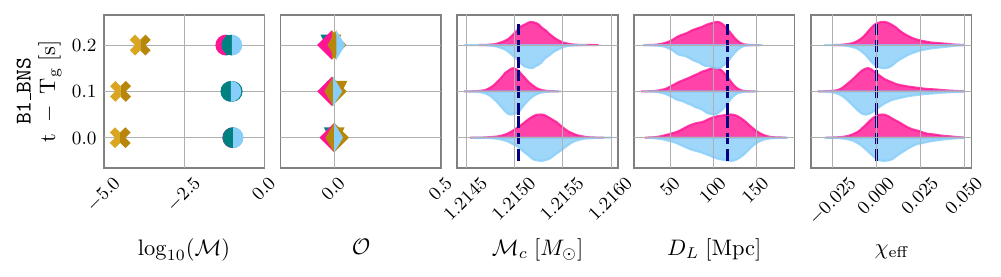}
	\caption{Results for CBC signals injected on top of blip glitches for high-mass BBH (top), low-mass BBH (middle), and BNS (bottom). Each row represents an instance of a glitch and a CBC signal injected at different times ($y$ axis) with respect to the glitch; compare run labels to Table~\ref{tab:settings}. The first two columns follow Fig.~\ref{fig:overlaps_example} and Table~\ref{tab:matches}. The violin plots show marginalized posteriors for select recovered parameters, specifically from left to right: detector-frame chirp mass ${\cal{M}}_c$, luminosity distance $D_L$, and effective spin $\chieff$. The pink (light blue) violin plots show the posteriors recovered with a ``CBC+Glitch" (``CBCOnly") analysis. The correct, injected value is plotted with a dashed, navy blue line. Pink and light blue are used for  $\cbcCBC$ and $\cbcG$ respectively, gold is used for glitch reconstructions, teal is used for CBC injections. The split colors indicate the two models used to calculate an overlap or mismatch.}
	\label{fig:Violin_Blip}
\end{figure*}

Blip glitches are short-duration glitches that occur in the most sensitive frequency band of the detectors with a frequency range from $\sim 32$Hz up through $1024$Hz. Because of their short duration and prevalence, they challenge analyses of high-mass events~\cite{Cabero_2019}, especially for sources with unequal masses and high spins~\cite{Ashton:2021tvz}. Their origin is largely unknown; $< 8 \%$ of LHO glitches and $ < 2 \%$ of LLO glitches were identified with auxiliary channels during the first and second observing runs~\cite{Cabero_2019}. Details about the blip glitches used for this study are provided in Table~\ref{tab:settings} while a spectrogram is provided in  Fig.~\ref{fig:Spectrograms_ALL}. By chance, the data around GPS time $1165578732$ contain a second glitch in LHO, 1s later. The presence of the additional glitch did not require any modification of the priors since it occurs entirely after the signal and has a low SNR. Figure~\ref{fig:Violin_Blip} presents our results, with runs labeled according to Table~\ref{tab:settings}. Each row corresponds to the same data with a given glitch and the same CBC signal injected at various times relative to the glitch. The merger time of the CBC relative to the glitch center is given on the $y$ axis.

The first and second column show various mismatches following the format of Fig.~\ref{fig:overlaps_example}. The set in the first column measure how well the models were reconstructed: the lower the mismatch, the more faithful each model recovery is. Specifically in all cases we find $\MgGgCBC \leq 0.01$, suggesting that the recovered glitch model does not consume any significant amount of the injected CBC signal, nor does it miss part of the glitch due to the presence of the signal. The first column also shows that typically $\McbcGinj\sim\McbcCBCinj$, though we also occasionally find $\McbcGinj<\McbcCBCinj$; in these cases the CBC model captures part of the glitch if we ignore the presence of the latter by analyzing the data assuming pure Gaussian noise.

In the second column we present information quantifying how similar the glitch the CBC are and, by extension, how difficult it is to separate them. Again the format is similar to Fig.~\ref{fig:overlaps_example}. Since $\OgGinj$ is evaluated on the injected CBC parameters and not maximized over CBC parameters, it does not directly correspond to how well a CBC template can recover the glitch, but it is a conservative estimate of the similarity between the glitch and injected signal. All overlaps are small, however these is some clear variation. Specifically, we find $ \OgGinj \sim \OgCBCgG\leq\OcbcCBCgG$  which means that in the ``CBCOnly" analysis the CBC model absorbs part of the glitch power. This is not the case with the full ``CBC+Glitch" model; indeed in all cases $\OgCBCgG$ is closer to the original value of $\OgGinj$.

The remaining columns show the marginalized posterior distributions for the (detector frame) chirp mass ${\cal{M}}_c$, luminosity distance $D_L$, and effective spin $\chieff$ respectively. The light blue posterior correspond to $\cbcCBC$ (``CBCOnly") whereas the magenta posteriors correspond to $\cbcG$ (``CBC+Glitch"). In all cases the parameters recovered with the full ``CBC+Glitch" model are consistent with the injected value. We find differences in the posteriors for the same CBC signal injected at different times with respect to the glitch. This is expected for two reasons. Firstly, the glitch and CBC posteriors are not completely uncorrelated, and hence the marginal CBC posterior will not be exactly the same as if there was no glitch. Secondly, each CBC is injected at slightly different times, and hence is subject to a different realization of the detector Gaussian noise. This distinction becomes more important for shorter signals, and indeed we find that the posteriors become more similar as the CBC signal duration increases from top to bottom.
Additionally, we find numerous instances where the ``CBCOnly" posteriors are significantly shifted and even inconsistent with the injected value. These cases are typically accompanied by  $\McbcGinj<\McbcCBCinj$ (first column) and/or $\OcbcCBCgG<\OgCBCgG$ (second column). Similar biases were reported in~\cite{Macas:2022afm} for extrinsic signal parameters computed in low latency.

Blip glitches are one of the most common glitch types and are very similar to high-mass BBHs. However, they are also one of the most straightforward glitch types to deal with due to their compactness in time and similarity to wavelets. The analyses presented here typically used the default \BayesWave glitch settings (apart from cases where there were additional glitches in the data beyond blips), and could be easily automated.

\subsection{Slow-Scattering Glitches}\label{sec:Scattering}

\begin{figure*}
	\includegraphics[]{overlap_legend.pdf}
	\includegraphics[]{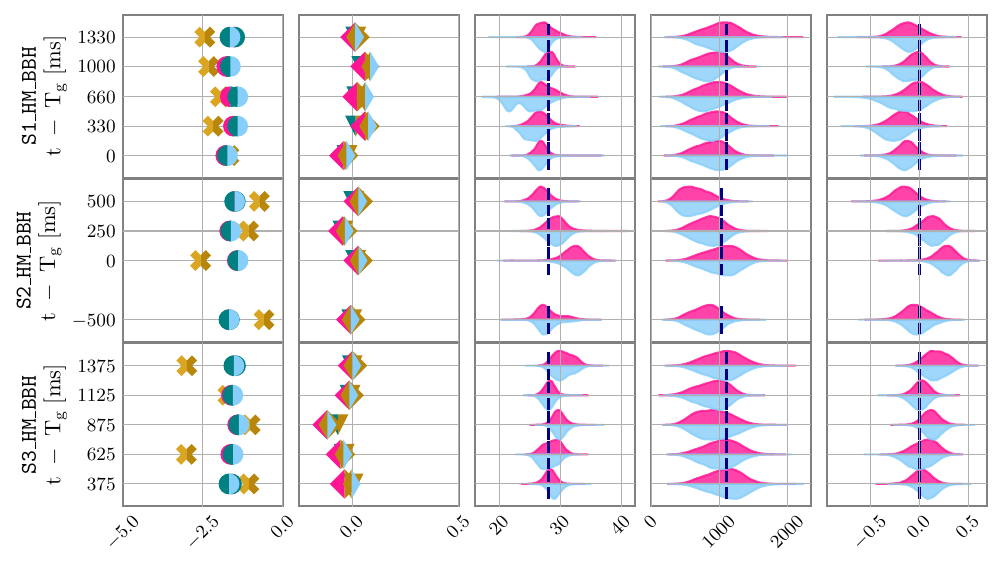}
	\includegraphics[]{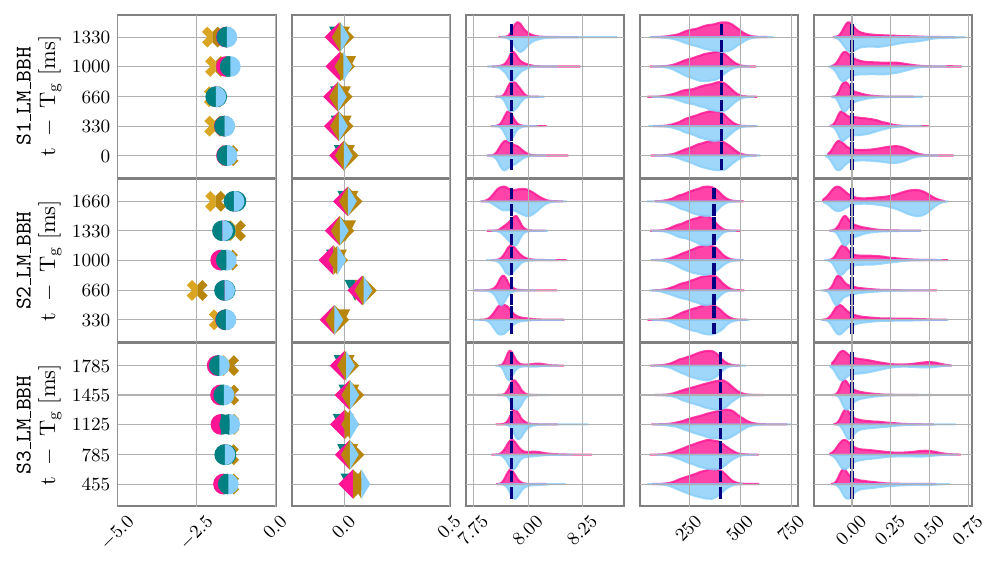}
	\includegraphics[]{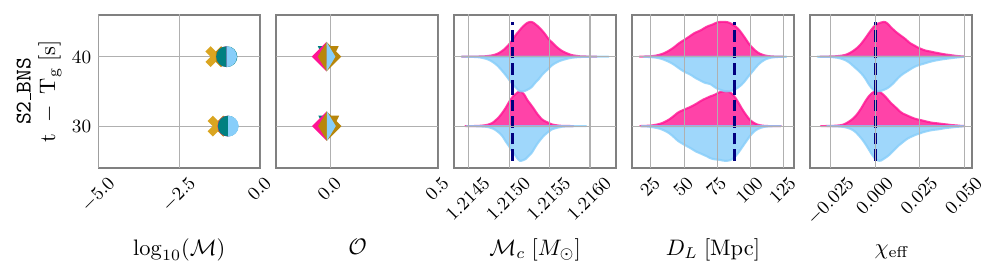}
	\caption{Same as Fig.~\ref{fig:Violin_Blip} but for the analyses anchored around slow scattering glitches. See Table~\ref{tab:settings} for run settings and labels. }
	\label{fig:Violin_Scattering}
\end{figure*}

\begin{figure*}
	\includegraphics[]{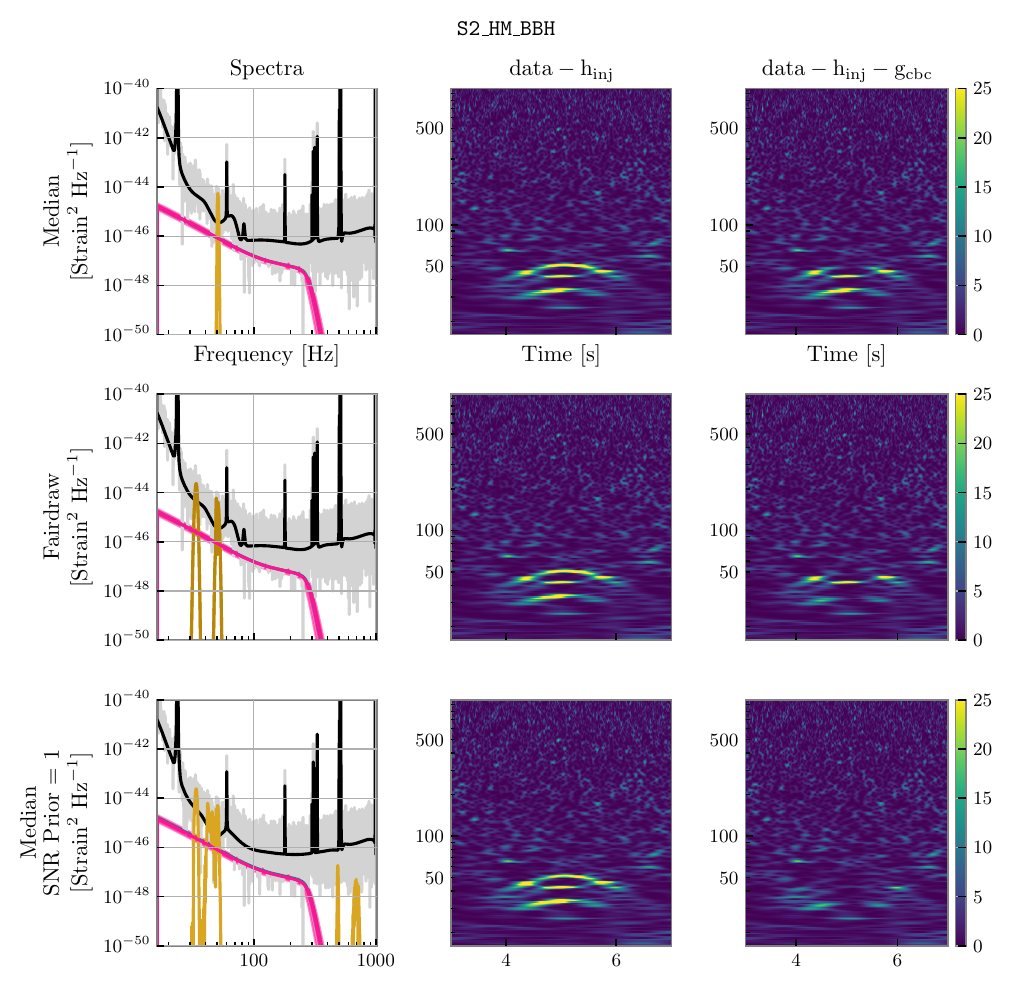}
	\caption{Spectra and residual plots for the {\tt S2\_HM\_BBH} injection at $-500$ms from the glitch, described in Sec.~\ref{sec:Scattering}. The first column shows the power spectra of $\inj$ (teal), the raw data (grey) and a point estimate for the PSD (black), $\cbcG$ (pink), and $\gCBC$ (gold). The second column shows spectrograms of the original data without the injection, while the third column shows spectrograms of the data where an estimate of the glitch has also been subtracted. The first and second row show results with the median glitch reconstruction and a fair draw from the posterior respectively. The third row shows results where the glitch amplitude prior set to peak at SNR=1 that aids identification of lower-SNR glitches. Due to the low SNR of the glitch, we find that all these choices impact the quality of glitch reconstruction and the subtraction. }
	\label{fig:Scattering_residuals}
\end{figure*}

Glitches from slow-scattered light appear in the detectors as long duration, $\mathcal{O}(4s)$, arches in the time-frequency domain, evenly stacked in frequency, usually in the range $8-64$ Hz. Each set of arches often recurs multiple times in the detector as shown in Fig. \ref{fig:Spectrograms_ALL}. Unlike blip glitches, they are not morphologically similar to CBC signals, yet they create long periods of non-Gaussianity and nonstationarity in the data, thus posing a challenge for noise PSD estimation. Their rate of occurrence increased during O3, when slow scattered light glitches overlapped with nine events throughout the observing run~\cite{gwtc3, O3a_catalogue}. Due to their morphology, slow scattered light glitches required longer analysis segments and wavelets of higher quality factors compared to short duration glitches such as blip glitches. A few of the glitches also extended to lower frequencies than other glitch types, so we used $\flow=8$ Hz for come cases. See Table~\ref{tab:settings} for run settings.

Figure~\ref{fig:Violin_Scattering} presents our results. All recovered CBC posteriors from the full ``CBC+Glitch" analysis are consistent with the injected values. Unlike the blip glitch case discussed in Sec.~\ref{sec:Blip}, the ``CBCOnly" analysis that ignores the presence of the glitch returns largely unbiased posteriors as well, exhibiting mostly small shifts. This is likely due to the fact that fast scattering glitches are morphologically very different than the types of CBC signals we consider.

The left column shows some variation between the median recovered glitch reconstructions. Though $\MgGgCBC$ remains mostly low and around $0.01$, it can reach $\sim0.1$ in some cases mostly for the second scatter glitch (runs whose label starts with {\tt S2}). We explore this further in Fig.~\ref{fig:Scattering_residuals} where we plot the spectrum of the data, signal, and glitch as well as spectrogram of the data for the {\tt S2\_HM\_BBH} injection at $-500$ms compared to the glitch. In each row, we plot the spectrum (left panel) and subtract from the data (right panel) the median glitch reconstruction (top panel) or a fair draw from the posterior (middle panel). In the middle right panel, more of the glitch has been subtracted compared to the top right panel. This is due to the low SNR of the glitch ($11.9$ for {\tt S2} compared to $15.1$ for {\tt S1}; computed by the Omicron pipeline~\cite{Robinet:2020lbf}), which results in some of the scattering arches residing in the threshold for reconstruction by the glitch model and thus not consistently included in the median. As a result, the median reconstruction has a large variation between different analyses, resulting in higher values for $\MgGgCBC$. In such cases, the glitch-subtracted data are sensitive to the choice of which glitch reconstruction to subtract (median or some fair draw) and additional case-by-case attention is needed. 

Motivated by the low SNR of the {\tt S2} glitch, we also considered the effect of the default priors in {\tt BayesWave}. The prior on the amplitude of the wavelets is broad, but peaks at SNR=5 per wavelet by default, see Fig. 5 of~\cite{Cornish:2020dwh}. The bottom panel of Fig.~\ref{fig:Scattering_residuals} shows results with a wavelet amplitude prior that peaks at SNR=1 per wavelet. Clearly more of the low-SNR glitch is subtracted. We leave further tests of the prior tunings such as this to future work. We conclude that although analysis of high-SNR slow scattering glitches can be potentially automated, lower-SNR instances will require some user attention.

\subsection{Fast Scattering Glitches}\label{sec:FScattering}

\begin{figure*}
	\includegraphics[]{overlap_legend.pdf}
	\includegraphics[]{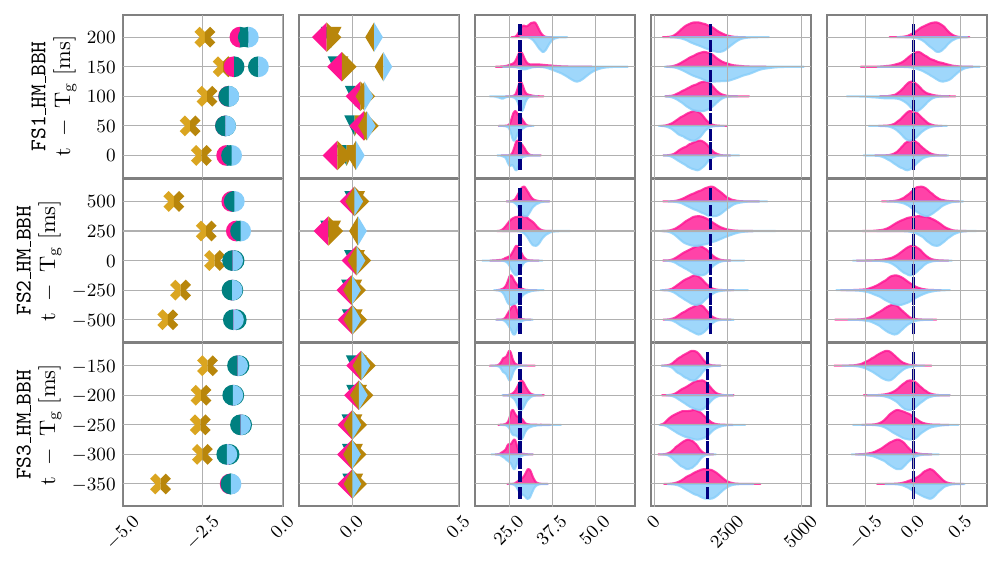}
	\includegraphics[]{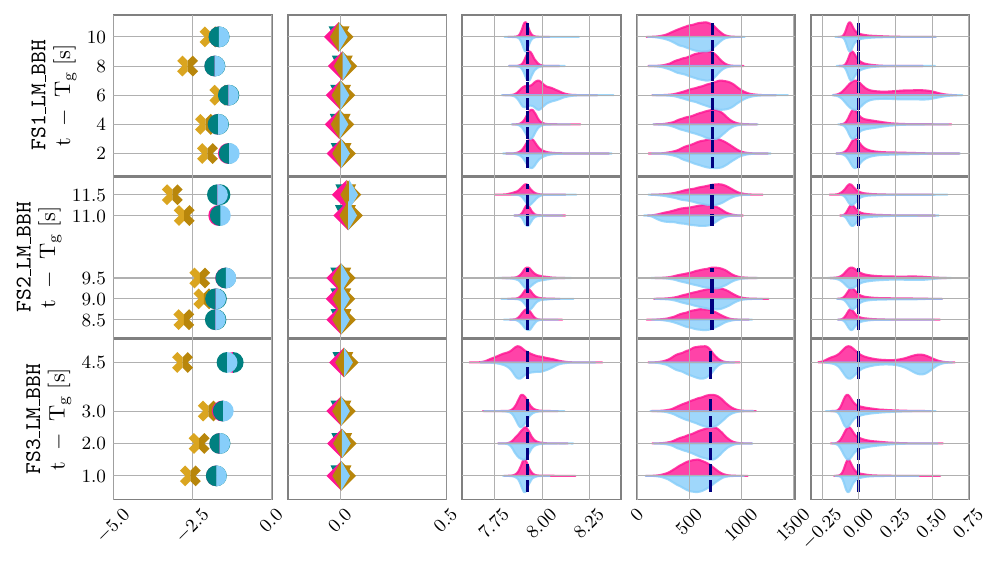}
	\includegraphics[]{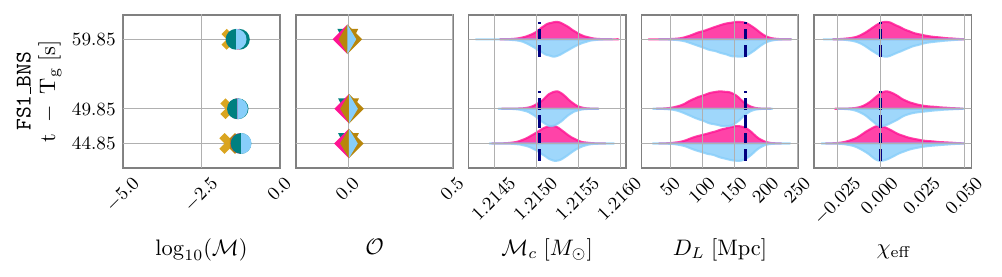}
	\caption{Same as Fig.~\ref{fig:Violin_Blip} but for the analyses anchored around fast-scattering glitches. See Table~\ref{tab:settings} for run settings and labels.}\label{fig:Violin_FastScattering}
\end{figure*}

Fast-scattering glitches (also referred to as ``crowns"~\cite{Soni:2021cjy}) are long duration glitches composed of many short bliplike bursts in frequencies from $~10-60$ Hz. Fast-scattering glitches have been linked to light scattered off the LIGO optical systems, particularly during ground motion~\cite{Soni:2021cjy}. They were the most common glitch type in LLO in O3, comprising 27\% of all glitches~\cite{Soni:2021cjy}. Two spectrograms of fast-scattering glitches are given in Fig.~\ref{fig:Spectrograms_ALL} and display the long- and short-term glitch behavior; relevant run settings are presented in Table~\ref{tab:settings}. 

Since a single fast-scattering glitch contains multiple, time-separated bursts, some adjusted settings are required. Such glitches create long-term nonstationarity, particularly at low frequencies so we increase the duration of the analyzed segments from 16 to 32 seconds for the low-mass BBHs. Despite the overall long duration of the glitch, we found that an increase in $\Qmax$ is not necessary, as each glitch consists of individual shorter burst that are each reconstructed by a few low-$Q$ wavelets.

Figure~\ref{fig:Violin_FastScattering} presents our results. Overall, we find similar results to the slow-scattering case of Fig.~\ref{fig:Violin_Scattering}, as the full ``CBC+Glitch" analysis is able to separate the signals and glitches while reliably estimating the parameters of the former. The ``CBCOnly" analysis returns mostly unbiased results, with the exception of isolated cases. Additionally, we find $\MgGgCBC<0.01$ in the first column which indicates that the glitch reconstruction is reliable even in the presence of a CBC signal. Due to the adjusted settings required for this analysis, automating analyses of fast-scattering glitches will be challenging.

\subsection{Tomte Glitches}\label{sec:Tomte}

\begin{figure*}
	\includegraphics[]{overlap_legend.pdf}
	\includegraphics[]{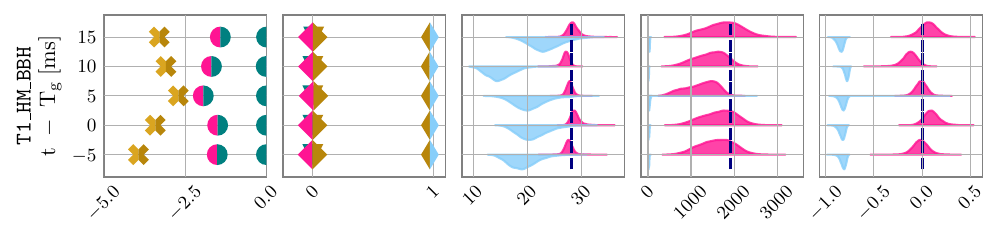}
	\includegraphics[]{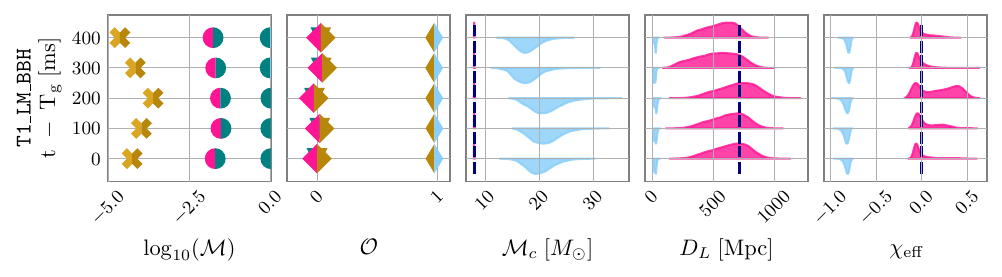}
	\caption{Same as Fig.~\ref{fig:Violin_Blip} but for the analyses anchored around a tomte glitch. See Table~\ref{tab:settings} for run settings and labels. The recovered CBC posteriors without simultaneous glitch modeling (light-blue violin plots) are heavily biased. }
	\label{fig:Violin_tomte}
\end{figure*}

\begin{figure*}
	\centering
	\includegraphics[]{overlap_legend.pdf}
	\includegraphics[]{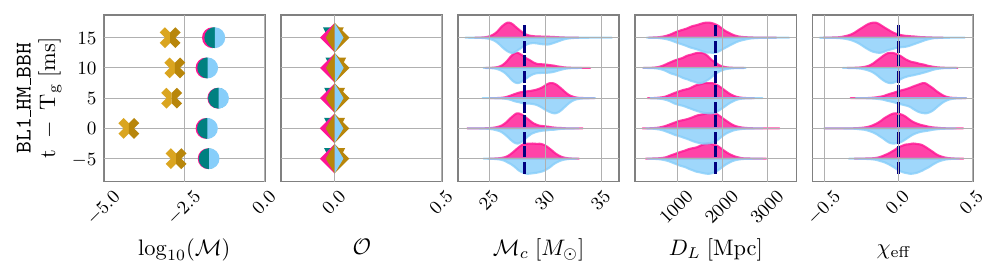}
	\caption{Same as Fig.~\ref{fig:Violin_Blip} but for the analyses anchored around a low-frequency blip glitch. See Table~\ref{tab:settings} for run settings and labels.}
	\label{fig:Violin_BlipLowFreq}
\end{figure*}

Tomte glitches are similar to blip glitches in that they can resemble CBCs with high, unequal masses and high spins~\cite{Ashton:2021tvz}. We again initially considered various instances of tomte glitches in LIGO data. However, we found the various tomte glitches to be morphologically very similar to each other, and therefore here restrict to a single instance. The glitch spectrogram is again given in Fig.~\ref{fig:Spectrograms_ALL} and results are presented in Fig.~\ref{fig:Violin_tomte}. We find broadly similar results to the blip glitch case.

Similar to blip glitches, the ``CBCOnly" analysis leads to large biases for all CBC parameters, both intrinsic and extrinsic. This suggests that of all glitches analyzed so far, tomtes are the ones most morphologically similar to CBCs. This is also evident in the second column of Fig.~\ref{fig:Violin_tomte}, where $\OcbcCBCgG \sim 1$ and $\OgGinj \sim 0$, which means that in the ``CBCOnly" analysis the CBC model ignored the signal entirely in favor of the glitch.  However, even in this challenging case, the joint ``CBC+Glitch" analysis is able to separate the signal and the glitch and result in reliable CBC parameter estimates and glitch reconstruction. 

\subsection{Low-Frequency Blip Glitches}\label{sec:BlipLowFreq}

\begin{figure*}
	\includegraphics[]{overlap_legend.pdf}
	\includegraphics[]{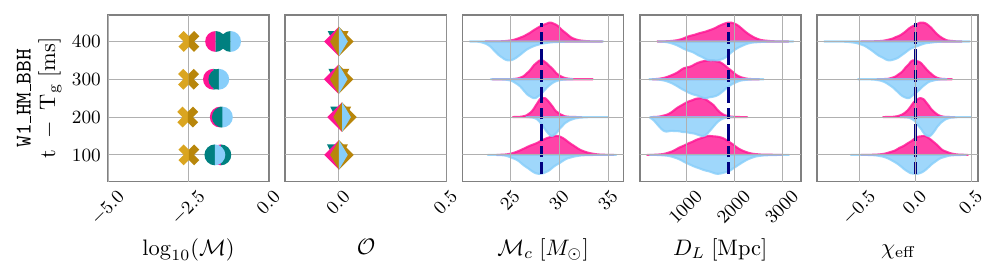}
	\caption{Same as Fig.~\ref{fig:Violin_Blip} but for the analyses anchored around the whistle glitch. See Table~\ref{tab:settings} for run settings and labels.}
	\label{fig:Violin_whistle}
\end{figure*}

\begin{figure*}
	\includegraphics[]{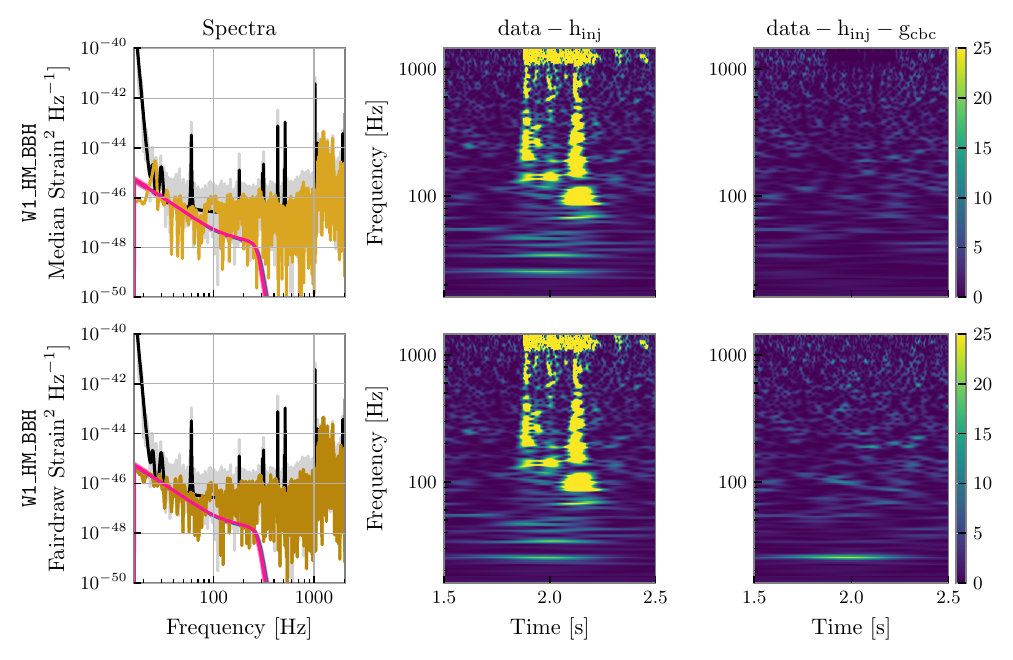}
	\caption{Spectra and residual plots for the {\tt W1\_HM\_BBH} injection at $100$ms from the glitch, described in Sec.~\ref{sec:Whistle} in similar format as Fig~\ref{fig:Scattering_residuals}. The right plot shows the final data where we have subtracted the median glitch reconstruction (top panel) or a fair draw from the glitch posterior (bottom panel), leaving behind Gaussian noise. The median reconstruction leads to oversubtraction of the glitch, we therefore favor the fair draw.}
	\label{fig:whistle_residual}
\end{figure*}

Low-frequency blips, as the name suggests, are similar in morphology to blip glitches except that they infect lower frequency bands, see the spectrogram in Fig.~\ref{fig:Spectrograms_ALL}. Given their similarity to blips, we consider a single instance of a low-frequency blip glitch and show results in Fig.~\ref{fig:Violin_BlipLowFreq}. We obtain similar results to the blip glitch case, Fig.~\ref{fig:Violin_Blip}, with small mismatches $\MgGgCBC\leq 0.01$ and recovered posteriors that are consistent with the injected values. However, low-frequency blips do not cause as significant a bias on the recovered CBC parameters when the glitch is not included in the model. This might be due to their low-frequency nature, which means that they do not significantly overlap with the CBCs in the most sensitive detector frequency band.

\subsection{Whistle}\label{sec:Whistle}

Whistle glitches are fairly loud glitches with a characteristic morphology depicted in the spectrogram of Fig.~\ref{fig:Spectrograms_ALL}. Our chosen instance of this glitch has an SNR of $\sim275$. Given their strength, more than $200$ wavelets are required to model them accurately, which poses a considerable challenge for sampler convergence. Given this fact, we only attempted injections on short duration segments. 

To aid convergence, we use {\tt GlitchBuster} to initialize the glitch model, see Sec.~\ref{sec:alg}. Despite the short duration, the high frequency of the glitch results in a lot of waveform cycles, we therefore also increase the maximum quality factor of the wavelets.
Finally, we also increase the number of iterations within the wavelet RJMCMC (see Fig.~\ref{fig:workflow}) from $10^2$ to $10^3$ as this glitch reconstruction requires upwards of 230 wavelets at every posterior sample. By default, we retain one out of 100 samples, and only update (i.e., add/remove/change) a single glitch wavelet at each sampler step. These default settings would therefore not lead to independent samples as not all wavelets have a chance to be updated before a posterior sample is retained. Details about the run settings are provided in Table~\ref{tab:settings}. 

 Results are presented in Fig.~\ref{fig:Violin_whistle} where we find that we are able to subtract the glitch consistently as well as estimate the CBC parameters. Despite the strength of the glitch, the ``CBCOnly" analysis returns mostly unbiased parameter posteriors, possibly due to the fact that whistle glitches are not morphologically similar to high-mass BBHs. The quality of glitch modeling and subtraction is further explored in Fig.~\ref{fig:whistle_residual} for the analysis of Fig.~\ref{fig:Violin_whistle}, specifically the injection at $100$ ms relative to the glitch. Comparison between the middle and right panel shows that we can efficiently subtract the glitch power. In this case, we also find that the median glitch reconstruction (top panel) results in an \emph{oversubtraction} of the glitch. The fair draw (bottom panel) leads to data that look more consistent with Gaussian noise. For this and the reasons discussed in Sec.~\ref{sec:Scattering} we generally prefer working with fair posterior draws rather than median glitch reconstructions when making glitch-subtracted data.

\subsection{Jensen-Shannon Divergence}
\label{sec:JSDivergence}

\begin{figure}
	\includegraphics[]{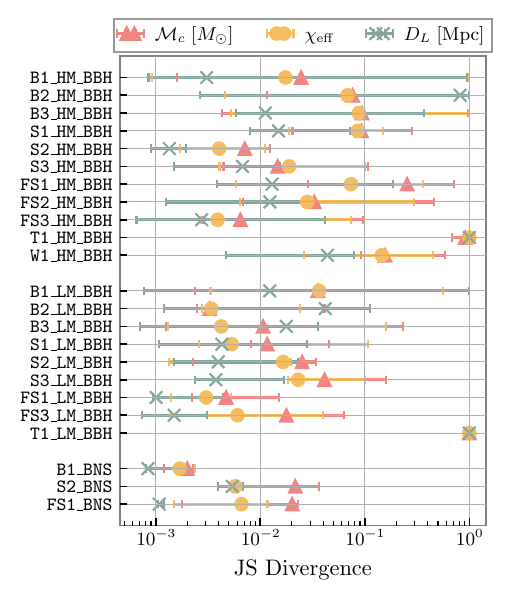}
	\caption{Jensen-Shannon (JS) divergence between the ``CBC only" and the ``CBC+Glitch" marginalized one-dimensional posteriors for $D_L$, $\chieff$, and ${\mathcal{M}}_c$.  We then plot the median (marker) and minimum to maximum values (error bars) over CBC injections at different times with respect to the same glitch. The breaks in the $y$ axis of the plot indicate different mass ranges, increasing upwards from the origin. The general trend is that JS increases with the signal mass, again suggesting that glitches affect high-mass systems more.}\label{fig:JSdivergence}
\end{figure}

As a final test, we compute a simple summary statistic for the differences between the ``CBCOnly" and the ``CBC+Glitch" posteriors: the Jensen-Shannon (JS) divergence.
The JS divergence describes the similarity of two distributions with JS = 0 for identical distributions and JS = 1 for disparate distributions. We plot the median and maximum/minimum JS for $D_L$, $\chieff$, and ${\mathcal{M}}_c$ across CBC injections on the same glitch at different times in Fig.~\ref{fig:JSdivergence}.  With the exception of the tomte glitch (discussed in Sec.~\ref{sec:Tomte}) where the posteriors are completely different, we recover the general trend that the JS divergence is smaller for low masses, which implies that the ``CBCOnly" and ``CBC+Glitch" posteriors are more similar for lower-mass events. This again supports the previous conclusion that though glitches are more likely to overlap long duration events, glitch subtraction is more important for high-mass than low-mass signals. Among the different parameters, the chirp mass is the one with the lowest JS on average, which is expected given the fact that it is the best measured intrinsic source parameter.

\section{Further validation studies}
\label{sec:futherStudies}

The results of Sec.~\ref{sec:glitches} show that the full ``CBC+Glitch" analysis can separate signals and glitches. In this section we provide some further validation tests regarding robustness against waveform systematics in the case of injections including higher-order modes or spin precession. We also assess the performance for signals that are observed by a single detector. Run settings and injection parameters for this section are presented in Tables~\ref{tab:Run_setting_further} and~\ref{tab:further_injtable} respectively. 

\begin{table*}
	\begin{tabular}{c|c@{\quad}c@{\quad}c@{\quad}ccccc}
		\hline
		Run Label   & GPS Time & Injected signal   & Glitches           &   $\Tobs$ [s] &   $\Tw$ [s] &   $\Qmax$ &   $\Dmax$ &   $\flow$ [Hz] \\
		\hline
		\tt{B1\_HM\_BBH\_HOM}  & 1168989748 & HM BBH w/ HOM  & Blip LHO   &      4 &    1 &     40 &    100 &     16 \\
		\hline
		\tt{T1\_HM\_BBH\_SPIN} & 1243679046 & HM BBH w/ SPIN & Tomte LLO  &      4 &    1 &     40 &    100 &      8 \\
		\hline
		\tt{B1\_HM\_BBH\_SING} & 1168989748 & HM BBH w/ SING & Blip LHO &      4 &    3 &     40 &    100 &  16 \\
		\hline
	\end{tabular}
	\caption{
		Settings for the runs of Sec.~\ref{sec:futherStudies} that test the effect of omitting higher-order modes (\texttt{HOM}), spin precession (\texttt{SPIN}), or considering only a single detector (\texttt{SING}). Columns give the same information as those of Table~\ref{tab:settings}.
	}
	\label{tab:Run_setting_further}
\end{table*}

\begin{table}
	\begin{tabular}{c|cccc}
		Signal & Injected Waveform & Varied param. & min. & max. \\
		\hline
		\texttt{HOM}  & \texttt{IMRPhenomHM} & $\cos(\iota)$  & -1       & 1   \\
		\hline
		\texttt{SPIN}  & \texttt{IMRPhenomPv2} & $\chi_p$ & 0.23 & 0.60 \\		
		\hline
	\end{tabular}
	\caption{Parameters of the injected signals for the tests of higher-order modes (\texttt{HOM}) and spin precession (\texttt{SPIN}). For high-order modes (spin precession) we vary $\cos(\iota)$ ($\chi_p$) between a minimum and a maximum value (third and fourth columns) to modify the strength of the deviation between the \texttt{IMRPhenomD} recovery waveform and the injected waveform. }\label{tab:further_injtable}
\end{table}

\subsection{Waveform Systematics: Higher-Order Modes}
\label{sec:HOM}

\begin{figure*}
	\centering
	\includegraphics[]{overlap_legend.pdf}
	\includegraphics[]{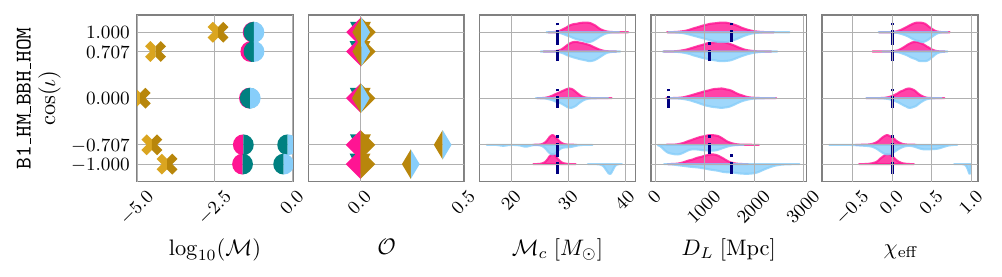}
	\caption{Similar to Fig.~\ref{fig:Violin_Blip} for a blip glitch and different high-mass BBH signals injected with higher-order modes but recovered without. The $y$ axis now shows the binary inclination. See Table~\ref{tab:Run_setting_further} for run settings and labels.}
	\label{fig:Blip_HOM_violin}
\end{figure*}

\begin{figure}
	\centering
	\includegraphics[]{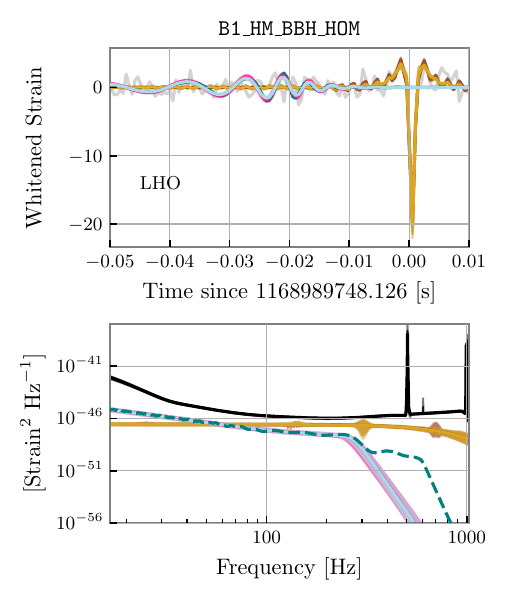}
	\caption{Whitened time-domain reconstruction (top) and spectrum (bottom) for a high-mass BBH injected with higher-order modes and edge-on. We show medians and 90\% credible intervals for the CBC signal (magenta), the glitch (gold), and the noise PSD (grey/black) from the ``CBC+glitch" analysis. The injected signal is given with a dashed teal line. The higher-order modes result in oscillations in the inspiral spectral amplitude as well as additional power at high frequencies. The CBC waveform used for the reconstruction does not include higher-order modes, however the reconstructed glitch model does not recover the excess power from higher-order modes. The CBC model from a ``CBCOnly" analysis (light blue) is similar to the one from the ``CBC+glitch" analysis.}
	\label{fig:Blip_HOM_reconstruction}
\end{figure}

The results of Sec.~\ref{sec:glitches} are based on {\tt IMRPhenomD}~\cite{Husa:2015iqa,Khan:2015jqa}, a waveform model that does not include higher-order modes, i.e., power from spherical harmonics beyond the dominant $l = |m| = 2$ mode. Such modes change the waveform morphology and thus neglecting them will lead to biases, especially for high SNR, unequal-mass systems, observed ``edge-on" ($\cos{\iota} = 0$)~\cite{Varma:2014jxa,CalderonBustillo:2016rlt,Pekowsky:2012sr,Varma:2016dnf,CalderonBustillo:2015lrt,Chatziioannou:2019dsz}\footnote{The inclination $\iota \in [0, \pi]$ is defined as the angle between our line-of-sight and the binary's Newtonian orbital angular momentum.}. Our current CBC sampler can work  with waveform models that include higher-order modes such as {\tt IMRPhenomHM}~\cite{London:2017bcn}, however, we do not perform such runs here because we lack an implementation of the heterodyne procedure~\cite{Cornish:2021wxy,Cornish:2021lje} that speeds up the likelihood calculation for such waveforms. Such an extension was described in~\cite{Leslie:2021ssu} and we plan to implement it in the future.

Because a real signal will inevitably contain some amount of higher-order modes, recovery with a waveform that neglects them could induce a systematic error in parameter extraction. Perhaps even more worrisome would the possibility that the glitch model subsumes some of the higher-order mode power which is then inadvertently subtracted from the data together with the glitch. We check for both effects by injecting the high-mass BBH signal from Table~\ref{tab:injtable} using {\tt IMRPhenomHM}~\cite{London:2017bcn} with varying inclination $\cos \iota\in [-1,1]$ onto one of our blip glitches and recover them again with {\tt IMRPhenomD}. Figure~\ref{fig:Blip_HOM_violin} shows recovered parameters for different system inclinations and Fig.~\ref{fig:Blip_HOM_reconstruction} shows the recovered CBC and glitch reconstruction for the case with $\cos \iota=0$.

Compared to the top row of Fig.~\ref{fig:Violin_Blip}, i.e., the same injected CBC and glitch but without higher-order modes and different inclinations, we find that $\MgGgCBC$ has increased, but still remains $\leq 0.01$. The other mismatches are comparable between Figs.~\ref{fig:Blip_HOM_violin} and~\ref{fig:Violin_Blip}. Despite the increase in $\MgGgCBC$, its value remains small and comparable to results from other blip glitches without higher-order modes, for example the second row of Fig.~\ref{fig:Violin_Blip}. We attribute this to the fact that the higher-order mode power is still too low to overcome the parsimony requirement of the glitch model to be picked up. 
Figure~\ref{fig:Blip_HOM_reconstruction} further reinforces this picture, by showing that the main effect of higher-order modes is additional high-frequency power that cannot be captured by the template (compare the magenta reconstruction to the teal injection). However, this residual power is still two orders of magnitude below the glitch power in the same frequency range.

Figure~\ref{fig:Blip_HOM_violin} also confirms that the amount of bias expected on the CBC parameters (notably $D_L$) is a function of the binary inclination. Furthermore, we find that not including the glitch in the model now leads to more pronounced parameter biases (blue violin plots). 
Regardless, even in the edge-on case the glitch and CBC signal can be separated sufficiently well as demonstrated by the mismatches in the first column.
Higher SNR signals or, in general, a signal with more than SNR$\sim 6-7$ in the higher-order modes could have more noticeable deviations from waveforms without higher-order modes where their power could then be picked up by the glitch model. However, current events with detectable higher-order mode content are below this SNR threshold~\cite{LIGOScientific:2020stg,LIGOScientific:2020zkf,Mills:2020thr,Zevin:2020gxf,Hoy:2021dqg}. 

\subsection{Waveform Systematics: Spin Precession}
\label{sec:spins}

\begin{figure*}
	\centering
	\includegraphics[]{overlap_legend.pdf}
	\includegraphics[]{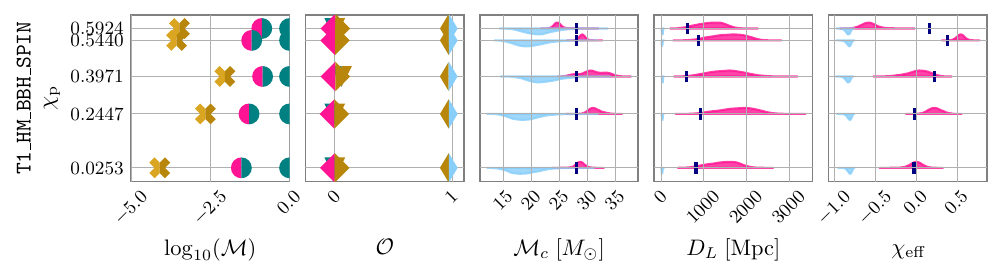} 
	\caption{Similar to Fig.~\ref{fig:Violin_Blip} but for a tomte glitch and different precessing high-mass BBH signals as a function of the injected binary precession parameter $\chip$. See Table~\ref{tab:Run_setting_further} for run settings and labels.}\label{fig:Tomte_chip_violin}
\end{figure*}

\begin{figure}
	\centering
	\includegraphics[]{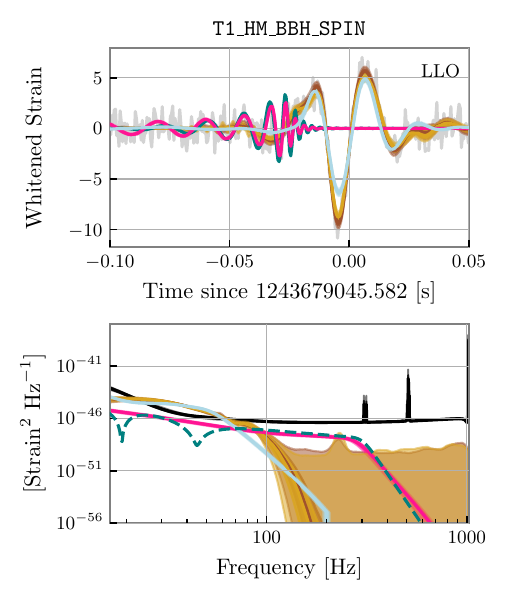} 
	\caption{Similar to Fig.~\ref{fig:Blip_HOM_reconstruction} for a spin-precessing high-mass BBH signal with $\chip = 0.592$. Spin precession induces oscillations in the inspiral spectral amplitude, most visible at around $40$ Hz. The CBC waveform template used for the reconstruction assumes the spins are aligned with the orbital angular momentum, however the reconstructed glitch model appears unaffected. The CBC model from a ``CBCOnly" analysis instead attempts to recover the glitch.  }\label{fig:Tomte_chip_reconstruction}
\end{figure}

In physical scenarios where the component spins are misaligned with the orbital angular momentum, the binary system experiences spin-precession which modulates the observed waveform~\cite{Apostolatos:1995pj}. The current implementation of the CBC sampler used here only accounts for spins aligned with the orbital angular momentum, so we assumed that the injected waveforms in our main analysis were also nonprecessing, motivated also by the lack of strong precession effects in event catalogues~\cite{gwtc3,LIGOScientific:2021psn}. However, signals with large in-plane spins, unequal masses, and/or observed edge-on could exhibit strong precessional effects~\cite{Apostolatos:1994mx,Schmidt:2010it,Schmidt:2012rh,Pratten:2020igi,Chatziioannou:2014bma,Kesden:2014sla,Chatziioannou:2016ezg,Khan:2018fmp,Pratten:2020ceb,Khan:2019kot,Ossokine:2020kjp}. The CBC sampler will be extended to include misaligned spin degrees of freedom in the future.

Similar to our analysis of the impact of higher-order modes, we study the impact of using nonprecessing templates by performing injections of high-mass BBHs with misaligned spins on the tomte glitch. The tomte glitch family was selected for this study as it is similar morphologically to highly-spinning, massive BBHs~\cite{Ashton:2021tvz}, and also because it consistently leads to the largest biases in CBC parameters when mismodeled, see Fig.~\ref{fig:Violin_tomte}. For the injections we use {\tt IMRPhenomPv2}~\cite{Hannam:2013oca} and we recover the signals with the same {\tt IMRPhenomD} waveform as before. The degree of spin-precession in a signal is commonly characterized by the parameter $\chip$ which is proportional to a mass-weighted maximum (over the two compact objects) of the in-plane spin magnitude~\cite{Schmidt:2014iyl}. Its range is $\chip \in \left[0, 1\right]$, where $\chip = 0$ describes a system with aligned spins (no precession) and $\chip = 1$ is a maximally precessing system. Figure~\ref{fig:Tomte_chip_violin} shows results for different values of $\chip$ and Fig.~\ref{fig:Tomte_chip_reconstruction} shows the recovered CBC and glitch reconstructions for the case the largest $\chip$. 

Compared to Fig.~\ref{fig:Violin_tomte} that shows results with the same signal and glitch but with a nonprecessing injection, we find that parameter recovery is increasingly biased as the injected $\chi_p$ increases. Again the ``CBCOnly" analysis (blue violin plots) leads to overwhelming parameter biases, which seems to be a generic feature of tomte glitches. This is again reflected in the fact that $\OcbcCBCgG \sim 1$ whereas $\OgGinj \sim 0$. Importantly, however, $\MgGgCBC \leq 0.01$ and is similar to the corresponding mismatches of Fig.~\ref{fig:Violin_tomte}; this means that the excess power due to spin-precession is not recovered by the glitch model, as its power is too low to be significant to the glitch model.

A similar conclusion is drawn from Fig.~\ref{fig:Tomte_chip_reconstruction}. The recovered signal is noticeably different from the injected signal, most prominently shown in the bottom panel where the precession-induced amplitude oscillations are absent from the posterior. However, again the glitch reconstruction appears to be unaffected, most likely because the relevant residual power is 2--3 orders of magnitude below the glitch power in the relevant frequency range.
Higher binary inclinations and higher signal SNRs might make the difference between precessing and nonprecessing waveforms more stark, causing the glitch model to capture any residual power due to spin precession. However, with current signal strengths and inferred amounts of spin precession, we find this to be unlikely. Though the CBC parameters recovered are clearly biased, the glitch modeling appears to be robust. Glitch-subtracted data can therefore be constructed and further analyzed with more complete waveform models.
Nonetheless, we plan to extend the analysis to include spin-precession effects in the CBC sampler.

\subsection{Single-detector signals}
\label{sec:multiIFO}

\begin{figure*}
	\centering
	\includegraphics[]{overlap_legend.pdf}
	\includegraphics[]{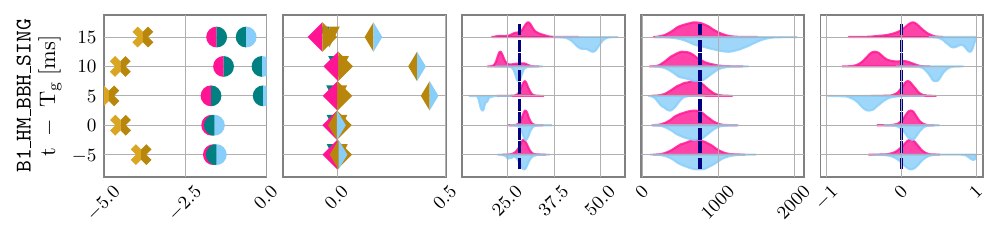} 
	\includegraphics[]{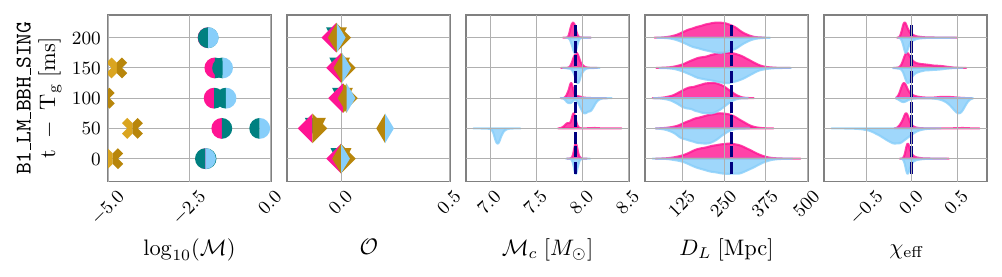} 
	\caption{Similar to Fig.~\ref{fig:Violin_Blip} but for a blip glitch and high-mass BBH (top) and low-mass BBH (bottom) signals observed in a single LHO detector. See Table~\ref{tab:Run_setting_further} for run settings and labels.}\label{fig:single_detector_violin}
\end{figure*}

\begin{figure}
	\includegraphics[]{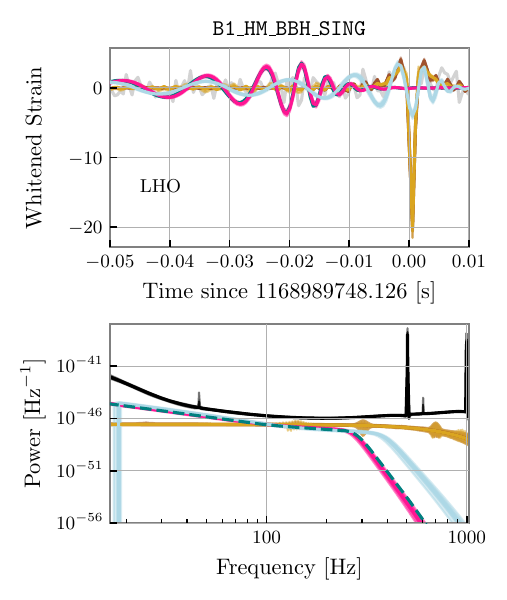} 
	\caption{Similar to Fig.~\ref{fig:Blip_HOM_reconstruction} for the {\tt B1\_HM\_BBH\_SING} injection at $100$ms from the glitch. When using the full ``CBC+Glitch" model,  $\cbcG$ and $\gCBC$ recover the CBC signal and glitch respectively even when data from a single detector only are available. The CBC model from a ``CBCOnly" analysis instead attempts to recover the glitch.  } \label{fig:single_detector_reconstruction}
\end{figure}

The joint analysis of CBCs and glitches assumes that the astrophysical signal is coherent across the detector network, while the glitch is not. Hence, the results presented so far are based on data from both LIGO detectors. However, single-detector candidates have been reported~\cite{gwtc3}, and we therefore test here if our analysis could separate them from glitches. The CBC waveform template we employ is a fairly restrictive model and we indeed find that this allows us in some cases to separate them from glitches even in single-detector data. Such a separation would be inherently impossible for the previous \BayesWave analyses that distinguished between signal and glitch solely via coherence within a detector network~\cite{Cornish:2020dwh}. 

We revisit runs \texttt{B1\_BBH\_HM} and \texttt{B1\_BBH\_LM} from Table~\ref{tab:settings} and analyze now only the LHO data that contain the glitch. We decrease the distance so that the single-detector SNR is $15$ for consistency with all other analyses. We present parameter results in Fig.~\ref{fig:single_detector_violin} and find that we are largely able to seperate the signal and the glitch. Even in this single-detector case the full ``CBC+Glitch" model outperforms the simpler ``CBCOnly" analysis and the glitch reconstruction is consistent with the one from data with no CBC injections. An example reconstruction plot is shown in Fig.~\ref{fig:single_detector_reconstruction} where again the CBC and glitch components of the full ``CBC+Glitch" model recover their corresponding data component. The ``CBCOnly" analysis, on the other hand, largely mistakes the glitch for a signal. Although these preliminary results are promising, we remain cautious of such cases. If presented with a similar scenario during an actual observing run, our analysis would require additional case-by-case attention and testing.

\section{Classifying triggers}
\label{sec:signalClassification}

\begin{table}
	\begin{tabular}{c|cccccccc}
		\hline
		Trigger   & GPS Time           &   $\Tobs$ [s] &   $\Tw$ [s] &   $\Qmax$ &   $\Dmax$ &   $\flow$ [Hz] \\
		\hline
		S191225 & 1261346253 &      4 &    1 &     40 &    100 &     16 \\
		\hline
		S200114 & 1263002916 &      4 &    2 &     40 &    100 &     16 \\
		\hline
		Tomte 1 & 1243679046 & 4 & 1 & 40 & 100 & 16 \\ 
		\hline
	\end{tabular}
	\caption{
		Settings for the analyses of Sec.~\ref{sec:signalClassification}. The first two columns provide the trigger name and GPS time. The remaining columns are the same as Table~\ref{tab:settings}. Additionally, S191225 used a sampling rate of $1024$ Hz whereas S200114 used a sampling rate of $2048$ Hz. 
	}
	\label{tab:Run_setting_real}
\end{table}

The joint ``CBC+Glitch" analysis simultaneously models CBC signals and glitches, however, the priors for both the CBC and the glitch model allow for the possibility of no CBC and/or no glitch in the data. In the glitch case, this is straightforward, as the model allows for $0$ wavelets in all detectors, as was for example recovered in the case of GW150914 in~\cite{Chatziioannou:2021ezd}. While the CBC priors ensure the presence of a CBC in the data, the luminosity distance prior extends to $10$Gpc, which effectively corresponds to a signal with negligibly small SNR. This suggests that our analysis can be used to assess whether certain detected excess power consists of a CBC signal, a glitch, both, or neither. We revisit two low-significance candidates from~\cite{O3a_catalogue,LIGOScientific:2021tfm} and analyze them with the joint ``CBC+Glitch" and the ``CBCOnly" analysis.  Analysis settings are provided in Table~\ref{tab:Run_setting_real}.

\subsubsection{S200114}
\label{subs:200114}

\begin{figure}
	\centering
	\includegraphics[width=0.99\linewidth]{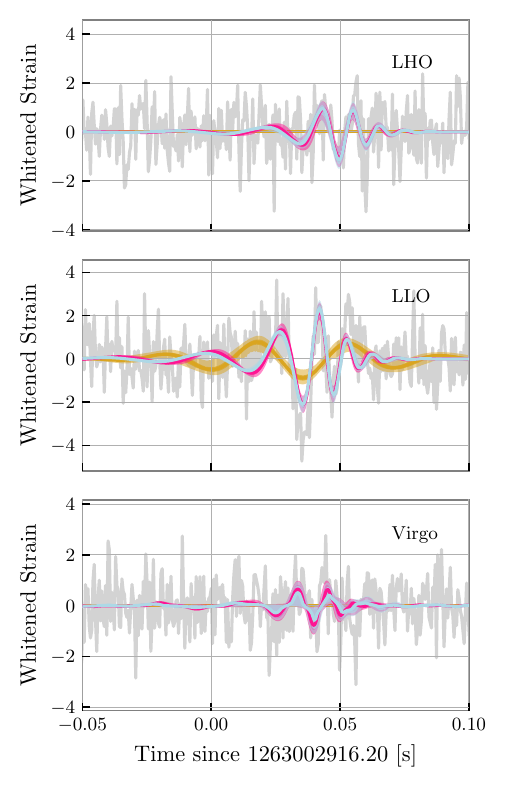}
	\caption{Whitened time-domain reconstruction for S200114 in each detector. We show medians and 90\% credible intervals for $\cbcG$ (magenta), $\gCBC$ (gold), and $\cbcCBC$ (blue). The data are consistent with the presence of both a CBC signal (i.e., coherent power that is morphologically similar to a CBC), and a glitch (i.e., additional incoherent power in LLO). }
	\label{fig:S200114_timedomain}
\end{figure}

\begin{figure}
	\centering
	\includegraphics[width=0.99\linewidth]{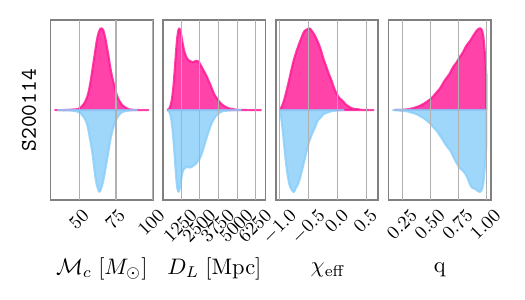}
	\caption{Posteriors for select parameters for S200114 from the ``CBC+Glitch" analysis (pink) to the ``CBCOnly" analysis (blue). Differences between these posteriors are smaller than those between different waveform models reported in~\cite{LIGOScientific:2021tfm}.}
	\label{fig:S200114_posterior}
\end{figure}

We begin with 200114\_020818 (referred to as S200114 from now on), which was identified by Coherent Wave Burst~\cite{Klimenko:2015ypf} with a false-alarm rate of 0.058 $\text{yr}^{-1}$~\cite{LIGOScientific:2021tfm}. Despite the low false alarm rate, the conclusion was that although an astrophysical origin could not be excluded, the trigger is of likely glitch origin since the estimated CBC parameters depended heavily on the choice of waveform model.

Figure~\ref{fig:S200114_timedomain} shows the time-domain reconstruction in each detector and Fig.~\ref{fig:S200114_posterior} shows a few marginalized parameter posteriors. The joint ``CBC+Glitch" analysis is consistent with the presence of both a CBC signal (at the 90\% credible level) and a low-frequency glitch in LLO. The morphology of the LLO glitch is consistent with a low-frequency blip or a tomte.
 This suggests that while there is excess power that is morphologically similar to a CBC and is coherent between LHO and LLO, there is also additional incoherent power in LLO that is captured by the glitch model. The CBC reconstruction is consistent with zero in Virgo at the 90\% credible level. Additionally, the ``CBCOnly" analysis finds a broadly consistent CBC reconstruction, with small differences at low frequencies. Differences between the ``CBCOnly" and ``CBC+glitch" CBC parameter posteriors are much smaller than the waveform systematics reported in~\cite{LIGOScientific:2021tfm}.

The presence of some amount of coherent power is consistent with the low false-alarm rate reported by a coherent detection pipeline in~\cite{LIGOScientific:2021tfm}.
The additional incoherent power in LLO could explain the inconsistent parameter estimation results between different waveform models presented in~\cite{LIGOScientific:2021tfm}, especially since the  $\tt{NRSur7dq4}$~\cite{Varma:2019csw}, $\tt{SEOBNRv4PHM}$~\cite{Ossokine:2020kjp}, and $\tt{IMRPhenomXPHM}$~\cite{Pratten:2020ceb} waveform models used in~\cite{LIGOScientific:2021tfm}  include the effects of spin-precession and higher-order modes. Thus they can account for more complicated morphologies~\cite{Pang:2018hjb,CalderonBustillo:2018zuq} in the CBC signal than the {\tt PhenomD} model used here. 

\begin{figure}
	\centering
	\includegraphics[width=0.99\linewidth]{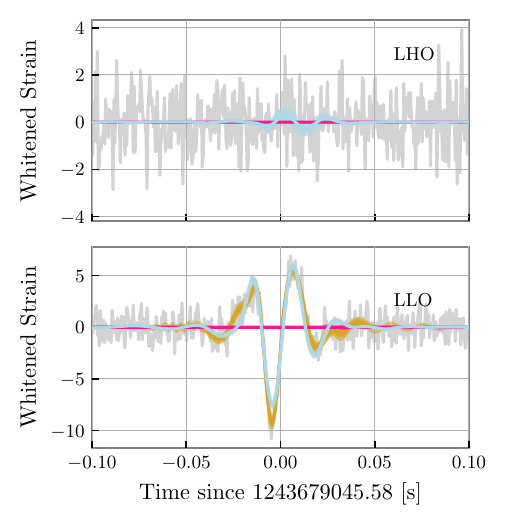}
	\caption{Whitened time-domain reconstruction for an analysis of data that contain only a single tomte glitch. We show medians and 90\% credible intervals for $\cbcG$ (magenta), $\gCBC$ (gold), and $\cbcCBC$ (blue). The data are consistent with the presence of solely a tomte glitch in LLO and do not recover any coherent power (the magenta CBC reconstruction is consistent with zero) when there is none. }
	\label{fig:Tomte_timedomain}
\end{figure}

The glitch in LLO is similar to a tomte which we have found to be recovered as a CBC in ``CBCOnly" analyses. To check whether a single tomte glitch could trick the ``CBC+glitch" analysis into concluding that both a CBC and a glitch are present in the data, we revisit the tomte glitch from Table~\ref{tab:settings}, perform no CBC injection, and carry out a ``CBC+Glitch" analysis. Reassuringly, the sampler indeed converges to the correct answer as shown in Fig.~\ref{fig:Tomte_timedomain}, namely that no CBC is present in the data, rather only a glitch. This suggests that the results of Fig.~\ref{fig:S200114_timedomain} cannot be the outcome of a single tomte glitch and the ``CBC+glitch" analysis does not recover coherent power when there is none.

However, we do not attempt to obtain a background estimate and thus cannot assess the probability that such a combination of coherent/incoherent power has a terrestrial origin. Such a full background estimate could result in the calculation of a false alarm rate similarly to the \BayesWave analysis in~\cite{LIGOScientific:2019ppi,KAGRA:2021tnv} based on the signal and glitch models only. In our case, we would use the ``CBC+glitch" analysis to estimate Bayes Factors for the various models of interest and compare them to similar results obtained from data that have been shifted in time between LHO and LLO.

\subsubsection{S191225}
\label{subs:S191225}

\begin{figure}
	\centering
	\includegraphics[width=\linewidth]{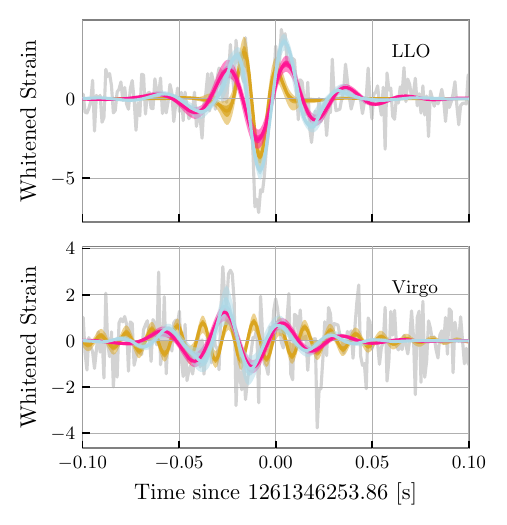}
	\caption{Whitened time-domain reconstruction for S191225 in LLO (top) and Virgo (bottom). We show the median and 90\% credible intervals for $\cbcG$ (magenta), $\gCBC$ (gold), and $\cbcCBC$ (blue). Some low-frequency coherent power is recovered by $\cbcG$ whereas the high-frequency power is largely incoherent and is recovered by the glitch model.  }
	\label{fig:S191225_timeseries}
\end{figure}

\begin{figure}
	\centering
	\includegraphics[width=0.99\linewidth]{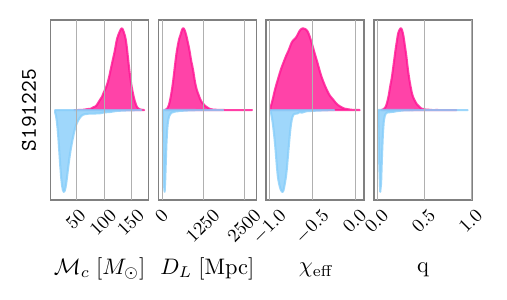}
	\caption{Posteriors for select parameters for S191225 from the ``CBC+Glitch" analysis (pink) to the ``CBCOnly" analysis (blue). }
	\label{fig:S191225_posterior}
\end{figure}

We then consider 191225\_215715 (labeled S191225 from now on), a low-significance LLO-Virgo trigger found by the PyCBC Live~\cite{Nitz:2018rgo} and the PyCBC-IMBH~\cite{Chandra:2021wbw} searches with false-alarm rates of 0.4 $\textrm{yr}^{-1}$ and 0.47 $\textrm{yr}^{-1}$ respectively in O3~\cite{gwtc3, LIGOScientific:2021tfm}. This candidate was ultimately deemed a glitch due to similar detector behavior surrounding the event. The reconstructed time-domain signal is shown in Fig.~\ref{fig:S191225_timeseries} for each detector, where we find that the data are consistent with a very high mass CBC and a glitch.  

When contrasting parameters from the ``CBCOnly" analysis to the ``CBC+Glitch" analysis. Fig.~\ref{fig:S191225_posterior}, we find that the former displays the telltale signs of a glitch; negative $\chieff$ and unequal masses~\cite{Ashton:2021tvz}. The latter still recovers some coherent power with recovered parameters instead pointing to a much higher mass binary (total mass $ > 300 M_{\odot}$). Since the new recovered mass is much larger than the original one, we might expect the false-alarm rate for this event to increase, although a full background estimate is outside the scope of this study. 

Overall, we find that S191225 is consistent with a glitch in LLO (possibly a tomte) atop of some coherent power, which agrees and expands upon with the conclusions of~\cite{gwtc3,LIGOScientific:2021tfm}.

\begin{figure}
	\includegraphics[width=0.99\linewidth]{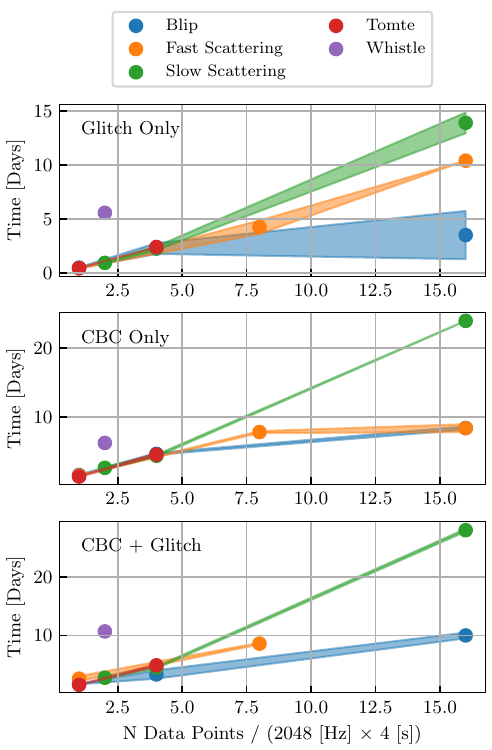}
	\caption{Run time estimates ($90\%$ intervals) for each glitch type and run setting we employ as a function of data points $N$ (segment length $\times$ sample rate). The x-axis is normalized by the shortest runs performed. Since runtime is (approximately) linear with the number of MCMC samples and number of chains, we rescale estimates to Number of Chains = $20$ and Number of Iterations = $4\times 10^6$, which are the default settings. Lighter settings can be used to expedite certain analyses. }\label{fig:runtimes}
\end{figure}

\section{Conclusions}
\label{sec:conclusions}

The various models that form the \BayesWave algorithm allow us to analyze GW data that include multiple components, specifically noise PSD, glitches, and a CBC signal. 
We present multiple tests with injected signals that overlap with real LIGO glitches and show that we can reliably separate signals and glitches, estimate the CBC parameters, and provide estimates for the glitch to be subtracted from the data. Runtime estimates for the various analyses are presented in App.~\ref{appendix:analysis}.

Our analysis is able to identify all glitches analyzed. It is particularly reliable for short-duration glitches such as blips, tomtes, and low-frequency blips which could be tackled in an automated way with default analysis settings. These are also the glitch types that cause the largest biases for CBC parameters when left unaccounted for.
Long-duration glitches such as fast and slow scattering glitches are more challenging and in some cases need specialized settings. Crucially, the necessary settings (such as maximum wavelet quality factor or analysis segment length) differ even between glitches of the same family, suggesting that automation is not yet feasible. 
This effect is particularly prominent for fast-scattering glitches that create long periods of nonstationarity that challenge PSD estimation, especially at low frequencies (below $40$ Hz).
Luckily, these glitches incur smaller biases in CBC parameter, most likely for the same reason they are difficult to model: they have a large time-frequency footprint that does not resemble a CBC chirp.

In this study, (with one exception) changes on glitch wavelet priors concerned their ranges, while their shapes were unaltered compared to default settings. However, dedicated glitch priors that target particularly problematic glitch families would improve glitch modeling. For example, a prior that favors wavelets with lower amplitude can lead to improved results for the low-SNR  slow-scattering glitch ({\tt S2}). Further examples of dedicated priors include a trained model based on a principal component analysis for tomte glitches~\cite{Merritt:2021xwh} or prior information about the frequency spacing of the scattering arches~\cite{Accadia_2010, 2020arXiv200714876S}. Depending on the characteristics of the most prevalent and problematic glitch types in O4, we plan to explore such dedicated priors in the future.


\acknowledgements

This research has made use of data, software and/or web tools obtained from the gravitational-wave Open Science Center (https://www.gw-openscience.org), a service of LIGO Laboratory, the LIGO Scientific Collaboration and the Virgo Collaboration.
Virgo is funded by the French Centre National de Recherche Scientifique (CNRS), the Italian Istituto Nazionale della Fisica Nucleare (INFN) and the Dutch Nikhef, with contributions by Polish and Hungarian institutes.
This material is based upon work supported by NSF's LIGO Laboratory which is a major facility fully funded by the National Science Foundation.
The authors are grateful for computational resources provided by the LIGO Laboratory and supported by NSF Grants PHY-0757058 and PHY-0823459.
S.H. and K.C. were supported by NSF Grant No. PHY-2110111.
D.D. was supported by the National Science Foundation as part of the LIGO laboratory
T.L. was supported by funding from the NASA LISA study office. 
N.J.C. was supported by NSF Grant No. PHY-1912053.
Software: gwpy~\cite{duncan_macleod_2020_3598469}, matplotlib~\cite{Hunter:2007}.


\appendix

\section{Runtimes}
\label{appendix:analysis}

Including a variable-dimensional glitch and PSD model comes at an additional computational cost compared to standard CBC analyses. We display the runtimes for the different glitch types and analyses presented in this paper in Fig.~\ref{fig:runtimes}. There is an essentially linear dependence between the number of data points (the segment length times the sampling rate) and the run time. The whistle glitch runs are outliers as additional settings were required to subtract such high-SNR glitches (see Sec.~\ref{sec:Whistle}). Since runtime is (approximately) linear with the number of parallel chains and the number of iterations, we rescale time by the default settings of 20 chains and $4\times 10^6$ iterations, though lighter settings can be used for expedited results. For example, the BNS runs for Fast-Scattering and Blip glitches used 10 chains for speed, and the whistle glitch was run with 25 chains for convergence. The rest were run with 20 chains. These estimates are further based on single-core runs and will be sped up accordingly by ongoing work to parallelize the chains.

\bibliography{Bib/OurRefs}
\end{document}